\documentclass[twocolumn,trackchanges]{aastex701}

\usepackage{amsmath}
\usepackage{bm}

\begin{document}
\title{Cosmic-ray electron propagation in NGC~3044 from radio continuum observations}

\author[orcid=0009-0003-9052-1976]{Shengtao Wang}
\affiliation{School of Physics and Astronomy, Yunnan University, Kunming 650500, China}
\email[show]{wstfch@mail.ynu.edu.cn}  

\author[orcid=0000-0002-3464-5128]{Xiaohui Sun} 
\affiliation{School of Physics and Astronomy, Yunnan University, Kunming 650500, China}
\email[show]{xhsun@ynu.edu.cn}

\author[orcid=0000-0002-2155-6054]{George Heald}
\affiliation{SKA Observatory, SKA-Low Science Operations Centre, 26 Dick Perry Avenue, Kensington, WA 6151, Australia}
\affiliation{CSIRO Space and Astronomy, PO Box 1130, Bentley, WA 6102, Australia}
\email{George.Heald@skao.int}

\author[orcid=0000-0001-6239-3821]{Jiangtao Li}
\affiliation{Purple Mountain Observatory, Chinese Academy of Sciences, 10 Yuanhua Road, Nanjing 210023, China}
\email{pandataotao@gmail.com}

\author[orcid=0000-0002-9390-9672]{Chao-Wei Tsai}
\affiliation{National Astronomical Observatories, Chinese Academy of Sciences, 20A Datun Road, Beijing 100101, China}
\affiliation{Institute for Frontiers in Astronomy and Astrophysics, Beijing Normal University, Beijing 102206, China}
\affiliation{School of Astronomy and Space Science, University of Chinese Academy of Sciences, Beijing 100049, China}
\email{cwtsai@nao.cas.cn}

\author{Judith Irwin}
\affiliation{Dept. of Physics, Engineering Physics \& Astronomy, Queen’s University, Kingston, K7L 3N6, Canada}
\email{irwinja@queensu.ca}

\author{Theresa Wiegert}
\affiliation{Instituto de Astrofísica de Andalucía (IAA-CSIC), Glorieta de la Astronomía s/n, 18008 Granada, Spain}
\email{twiegert@iaa.es}

\author[orcid=0000-0003-1778-5580]{Jun Xu}
\affiliation{National Astronomical Observatories, Chinese Academy of Sciences, 20A Datun Road, Beijing 100101, China}
\affiliation{National Key Laboratory for Radio Astronomy, Beijing 100101, China}
\email{xujun@nao.cas.cn}
 
\begin{abstract}

Star-forming edge-on galaxies often exhibit extended halo radiation in multiple bands, providing ideal laboratories for studying the transfer of matter from the disk to the halo. We investigate the transport of cosmic-ray electrons (CREs) and the associated galactic wind, and assess their impact on the surrounding medium in NGC~3044. We obtained the NGC~3044 total intensity image at 943~MHz from the Australian SKA Pathfinder (ASKAP) observations with a resolution of $16\arcsec$ and an rms noise of 20~$\mu$Jy~beam$^{-1}$. The sensitivity is higher than the previous observations at similar frequencies. We find that the ASKAP intensity profiles perpendicular to the disk can be fit with two exponential components. The scale heights of the thin and thick disks are $0.43\pm0.13$~kpc and $1.91\pm0.26$~kpc, respectively. By jointly fitting the total intensity and spectral index profiles with one-dimensional advection and diffusion models, we find that CREs are advected outward from the disk with the velocity increasing with height in a power law. Beyond $\sim3$~kpc, the velocity exceeds the escape speed of $\sim400\,\rm km\,s^{-1}$, indicating a strong wind. We further identify a possible superbubble of radius $\sim3$~kpc filled with soft X-ray emitting hot gas and surrounded by an H\textsc{i} shell and a bright H$\alpha$ rim. These results demonstrate that radio continuum observations provide a powerful probe of cosmic-ray–driven winds in normal star-forming spiral galaxies.

\end{abstract}

\keywords{\uat{Radio continuum emission}{1340} ---\uat{Cosmic rays}{329} ---\uat{Magnetic fields}{994} ---\uat{Spiral galaxies}{1560} ---\uat{Interstellar medium}{847}}

\section{Introduction} 
\label{sec:intro}

Galactic winds are multiphase and comprise cold molecular gas, dust, H\textsc{i}, warm ionized gas, hot X-ray emitting plasma, relativistic particles, and magnetic fields~\citep{veilleux2005,thompson2024}. They regulate star formation by removing disk gas, enrich the intergalactic medium with metals, and export magnetic energy beyond galactic boundaries. Cosmic-ray (CR) pressure is recognized as an efficient driver of the winds. CRs scatter on self-generated Alfvén waves and impart momentum and energy that establish vertical pressure gradients capable of lifting gas into the halo~\citep{everett2008}. Unlike purely thermal winds, CR-driven outflows can operate at lower star formation rates and sustain higher mass loads over larger radii~\citep{girichidis2018}.

NGC~3044 is an isolated, nearly edge-on barred spiral (SBc) galaxy with an inclination of $i\approx85\degr$, at a distance of $\sim20.3\,\rm Mpc$~\citep{irwin2013b,lee1997}. The galaxy is classified as non-starburst, with a star formation rate (SFR) of 1.75~$\rm M_{\odot}\,yr^{-1}$~\citep{vargas2019} and an SFR surface density of $3.7\times10^{-3}\,\rm M_{\odot}\,yr^{-1}\,kpc^{-2}$~\citep{wiegert2015}, but quite close to starburst~\citep{li2016}. \citet{tullmann2006a} argued for a disk-wide starburst rather than a nuclear one and found no evidence of an active galactic nucleus (AGN). NGC~3044 exhibits extraplanar emission in multiple ISM phases, indicating a disk–halo system driven by feedback~\citep{hummel1989,condon1990,colbert1996,lee1997,collins2000,tullmann2000,miller2003,tullmann2006a,lu2023}. H\textsc{i} mapping reveals a lopsided disk with numerous high-latitude features and large supershell signposts of energetic feedback disturbing the outer disk and seeding extraplanar gas \citep{lee1997,lee1999,zheng2022}. Deep H$\alpha$ imaging shows a bright extraplanar diffuse ionized gas (eDIG) layer extending to a height of $\sim1-2$~kpc with filaments and plumes, consistent with comparative DIG surveys \citep{rossa2000,vargas2019,lu2023}. XMM-Newton detects an extended soft X-ray halo that spatially links hot coronal gas to sites of star formation and eDIG~\citep{tullmann2006a,tullmann2006b}. Mid-IR Spitzer spectroscopy identifies halo polycyclic aromatic hydrocarbon features and H\textsc{ii} emission, demonstrating that dust and warm molecular gas participate in vertical flow \citep{rand2011}. 

Radio continuum observations show a prominent extended radio halo in NGC~3044, first reported by~\citet{hummel1989}. The GMRT 617~MHz imaging revealed distinct high-latitude features~\citep{irwin2013b}, but lacks the surface brightness sensitivity to trace the faint halo. With the Continuum Halos in Nearby Galaxies - an EVLA Survey (CHANG-ES) project at L- and C-band, \citet{irwin2024} confirmed a multiphase outflow and highlighted vertical features reaching several kpc. Polarimetric CHANG-ES analysis further shows coherent X-shaped halo magnetic fields in the sample, with NGC~3044 exhibiting strong vertical magnetic structures (“giant magnetic ropes”), indicative of magnetized galactic winds shaping the halo field~\citep{krause2020}. 

Complementary 1D CR-transport modeling (diffusion/advection) has already been applied to NGC~3044, yielding an approximately constant advection speed of $\sim200~\mathrm{km\,s^{-1}}$ from two-frequency (1.5/6~GHz) profiles \citep{heesen_2018b}. Building on this, we incorporate new low-GHz imaging at 943~MHz together with 6~GHz at matched resolution to strengthen the transport constraints. The Australian SKA Pathfinder (ASKAP), with its high resolution, high sensitivity, and wide-field capabilities at 943~MHz provides us with the opportunity to study the radio halo of NGC~3044.

The 943~MHz map recovers faint, extended halo emission and thus improves measurements of the thick disk scale heights and their uncertainties. Extending the frequency baseline from 943~MHz to 6~GHz provides direct leverage on the expected frequency dependence of the CR electron (CRE) propagation length $l_{\rm adv}\!\propto\!\nu^{-1/2}$ for advection and $l_{\rm diff}\!\propto\!\nu^{-1/4}$ for diffusion~\citep{krause2018}, making it easier to distinguish a steeper advection-like trend from a shallower diffusion-like one and to trace the wind to higher $z$ and to quantify its impact on the surrounding medium.

This paper is organized as follows: data acquisition and reprocessing are described in Sect.~\ref{sec:ob_data}; results are presented in Sect.~\ref{sec:results}; discussions are made in Sect.~\ref{sec:discussions}; conclusions are drawn in Sect.~\ref{sec:conclu}.

\section{Data acquisition and reprocessing}
\label{sec:ob_data}
\subsection{ASKAP data}

NGC~3044 was covered in the field observed by the ASKAP Pilot Survey for Rubin Deep Drilling Fields project (Project ID: AS310). To date, the galaxy has been observed twice, with observation IDs SB67559 and SB67758, and an integration time of 9 hours for each observation. The central frequency is 943~MHz, and the total bandwidth is 288~MHz, splitting into 1-MHz channels. 

\begin{figure}	
    \centering
    \includegraphics[width=0.98\columnwidth]{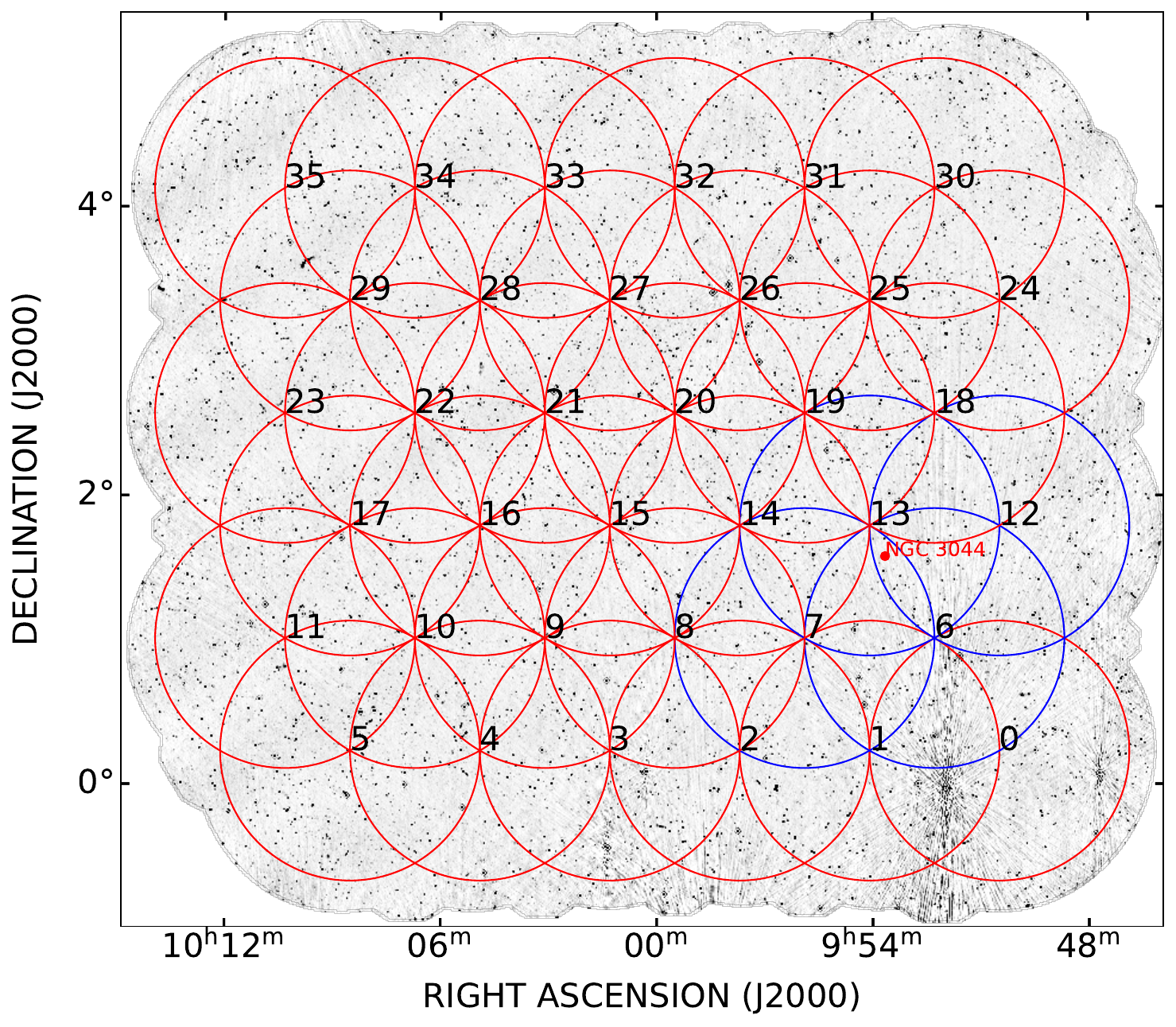}
    \caption{Layout of the 36 ASKAP beams with circles in the ``closepack36'' footprint configuration, overlaid on the $I$ image from CASDA. The radius of the circles is $0.9\degr$, which is approximately the primary beam width. The blue circles indicate the beams that cover NGC~3044.}
\label{fig:36beam_cover}
\end{figure}

Each ASKAP antenna is equipped with a phased array feed that forms 36 dual-polarization beams. The footprints of the beams are in the ``closepack36'' configuration as shown in Fig.~\ref{fig:36beam_cover}. NGC~3044 was fully covered by the four beams: 6, 7, 12, and 13, outlined by blue circles in Fig.~\ref{fig:36beam_cover}. 

The calibrated visibilities for each individual beam and the total intensity ($I$) images for the whole field, processed with {\small ASKAPsoft}\footnote{ASKAPsoft is the suite of processing software developed by the ASKAP computing team to process ASKAP observations.}, are public and available from the CSIRO ASKAP Science Data Archive \textit \rm {(CASDA)}\footnote{\url{https://data.csiro.au/domain/casdaObservation}}. 

We reprocessed the visibility data for beams 6, 7, 12, and 13 from CASDA using the pipeline SASKAP\footnote{\url{https://gitlab.com/Sunmish/saskap/-/tree/petrichor}} to improve calibration and imaging. 

For calibration, we performed two rounds of phase-only (p) and one round of amplitude-phase (ap) self-calibration. The solution interval was progressively reduced during each round, starting from 300~s at the beginning to 60~s in the end. 

For imaging, the fast generic widefield imager \textit{\rm {\small \textsc{WSclean}}}\footnote{\url{https://wsclean.readthedocs.io/en/latest/changelogs/v3.2.html}}~\citep{offringa2014,offringa2017} was used to generate a multi-frequency and multiscale synthesis image. To ensure that the ASKAP images are sensitive to the same range of spatial scales as the VLA observations, we restricted the UV coverage to $540-21,300\,\lambda$. This corresponds to a largest recoverable angular scale of $\sim6\arcmin$, which exceeds the $\sim4\arcmin$ size of NGC~3044; thus, no large-scale flux is missed. Briggs weighting~\citep{briggs1995} with \texttt{robust} = 0.25 was employed to enhance sensitivity towards faint, extended emission. Wgridder was employed to improve the accuracy of wide-field image reconstruction and to efficiently account for non-coplanar baseline effects~\citep{ye2022,arras2021}.

Primary beam correction is required to determine flux densities and combine images of different beams. Following the method by \citet{duchesne2024}, we built a robust model of the primary beam. We cross-matched the point sources in each ASKAP beam with the Rapid ASKAP Continuum Survey low-frequency \textit \rm {(RACS-Low)}\footnote{\url{https://data.csiro.au/collection/csiro\%3A52217v3}}~\citep{mcconnell2020,hale2021}, and obtained the ratio of flux densities as a function of the relative position to the center of each beam. 
A two-dimensional elliptic Gaussian was fit to the data to obtain the primary beam model, which was used to correct the images.

NGC~3044 is located closer to the center of beam 13 and lies near the edges of the other three beams (Fig.~\ref{fig:36beam_cover}). Therefore, we performed a weighted linear combination of the data only from beam 13 for the two observations, as these have the lowest noise and are almost free from sidelobe contamination. 
 
\subsection{VLA data}
From the \textit \rm {CHANG-ES}\footnote{\url{https://projects.canfar.net/changes/}}
 project~\citep{irwin2012a,irwin2024chang}, we obtained fully calibrated VLA L-band (1.57~GHz) C-array and C-band (6~GHz) D-array visibility data. The calibration was performed with the Common Astronomy Software Application (CASA)~\citep{mcmullin2007,casa2022} and is described in detail in previous CHANG-ES papers~\citep{irwin2012b,irwin2013,irwin2019,wiegert2015}. Each dataset was subsequently self-calibrated, except in cases where self-calibration did not yield a significant improvement.

We re-imaged the data with \textsc{WSclean}, again restricting the UV range to $540-21,300\,\lambda$ and adopting briggs weighting with \texttt{robust} = 0.25. To enhance our sensitivity to large-scale emission, we set \texttt{-multiscale-scale-bias} = 0.4 (smaller values place more weight on larger scale structure). The \texttt{-apply-primary-beam} option was used to correct for the primary beam response.

\begin{figure*}[!htbp]
    \centering
    \includegraphics[width=0.98\textwidth]{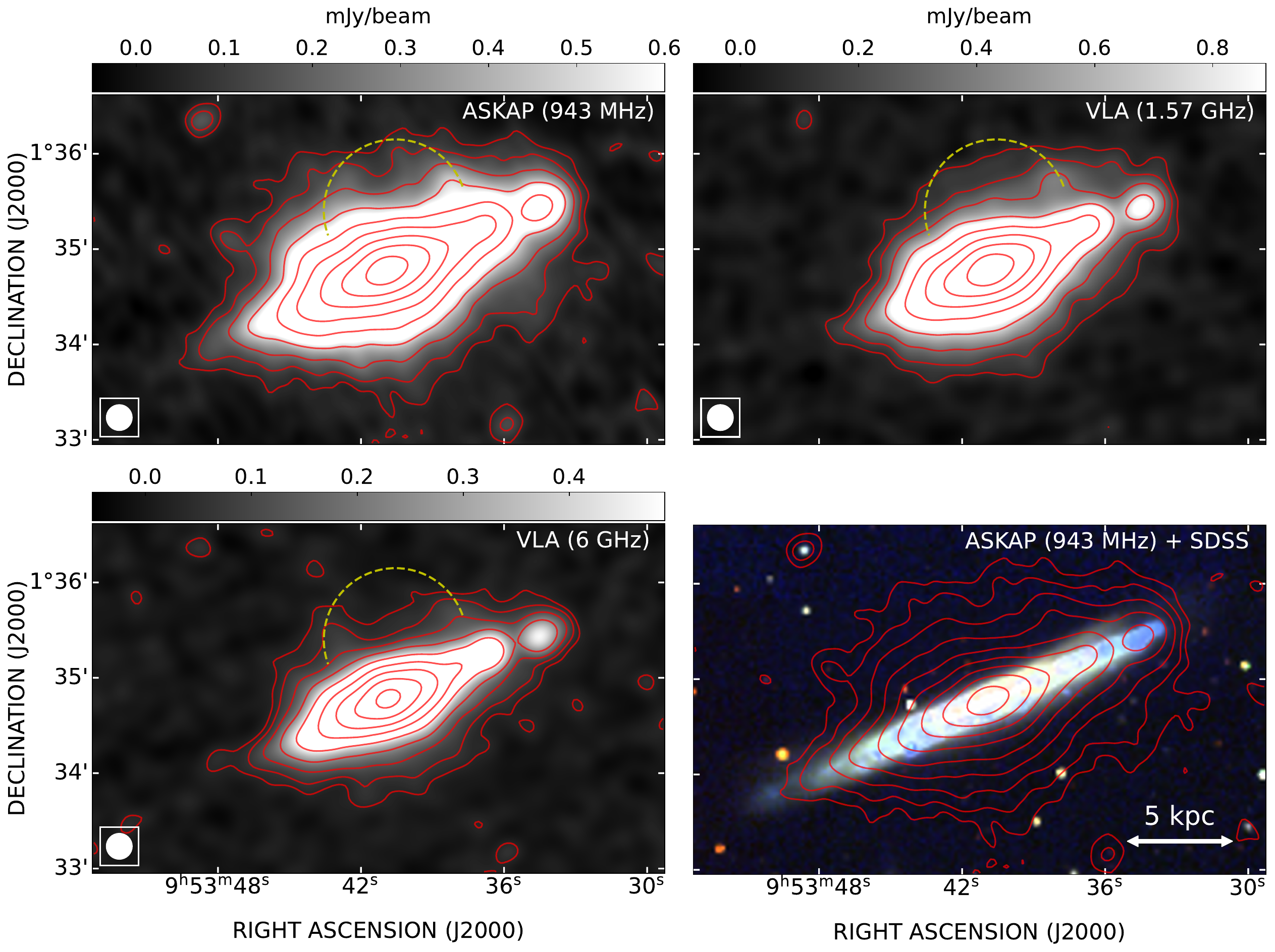}
    \caption{Total intensity images from ASKAP at 943~MHz with an rms noise of $\rm 20\, \mu Jy\,beam^{-1}$ (top left panel), from VLA at 1.57~GHz with an rms noise of $\rm 30\, \mu Jy\,beam^{-1}$ (top right panel) and at 6~GHz with an rms noise of $\rm 10\,\mu Jy\,beam^{-1}$ (bottom left panel). The VLA images have been convolved to the ASKAP resolution of $16\arcsec$. The ASKAP total intensity contours overlaid on an SDSS u, g, and r three-color image is shown in the bottom right panel. The Contours levels are at $3\sigma\times2^n$ ($n=0,\,1,\,2,\,\ldots$). The region of the superbubble is outlined by the yellow dashed semicircle.} 
\label{fig:total_intensity}
\end{figure*}

\section{Results}
\label{sec:results}

\subsection{Total intensity images and integrated flux densities}

 In Fig.~\ref{fig:total_intensity}, the total intensity map from ASKAP is shown in the top-left panel, with a resolution of $16\arcsec$ and an rms noise of 20~$\rm \mu Jy/beam$. In this map, we identify a possible superbubble. The VLA images in L-band and C-band are convolved to $16\arcsec$ and shown in the upper right and bottom left panels, with rms noises of 30 and 10~$\rm \mu Jy/beam$, respectively. The rms noise was measured in background regions near NGC~3044 but without emission from NGC~3044 and bright sources. NGC~3044 is either at or close to the beam center and has an angular size much smaller than the primary beam width, and the rms noise thus expects to be uniform across the source. Primary beam correction leads to an increase in noise toward the beam edges, which does not affect our estimate. The bottom right panel presents ASKAP total intensity contours overlaid on an optical background. It can be clearly seen that the radio emission extends far beyond the optical emission disk. Compared to the VLA images, the ASKAP image reveals halo emission much further out, to a height of $\sim7.7$~kpc above the mid-plane. This demonstrates the necessity of low-frequency radio continuum observations to trace CREs in the halo. 

From the total intensity images, we derived integrated flux densities in larger than 3$\sigma$ regions at 943~MHz, 1.57~GHz, and 6~GHz, assuming a relative calibration error of 5\% for both the VLA and ASKAP. We also collected measurements at low frequencies from 76~MHz to 227~MHz~\citep{Hurley-Walker+17} with the Murchison Widefield Array (MWA) from the GaLactic and Extragalactic All-sky MWA survey (GLEAM), assuming a 10\% relative calibration error. Other published measurements by GMRT and VLA were also collected. All flux densities $S_\nu$ at frequencies $\nu$ are listed in Table~\ref{tab:int_flux}.
 
\begin{deluxetable*}{cccccc}
\tablecaption{Integrated flux densities of NGC~3044.\label{tab:int_flux}}
\tablehead{
\colhead{Telescope} & \colhead{$\nu$} & \colhead{$S_\nu$} & \colhead{$f_{\rm th}$} & \colhead{Ref.} \\
\colhead{} & \colhead{(MHz)} & \colhead{(mJy)} & \colhead{(\%)} & \colhead{}
}
\startdata
MWA  & 76  & $432\pm 43$ & -- & \tablenotemark{a}   \\
MWA  & 84  & $433\pm 43$ & -- & \tablenotemark{a}   \\
MWA  & 92  & $467\pm 47$ & -- & \tablenotemark{a}  \\
MWA  & 99  & $512\pm 51$ & -- & \tablenotemark{a}  \\
MWA  & 107 & $495\pm 50$ & -- & \tablenotemark{a}  \\
MWA  & 115 & $465\pm 47$ & -- & \tablenotemark{a}  \\
MWA  & 122 & $405\pm 41$ & -- & \tablenotemark{a}  \\
MWA  & 130 & $381\pm 38$ & -- & \tablenotemark{a}  \\
MWA  & 143 & $399\pm 40$ & -- & \tablenotemark{a}  \\
MWA  & 151 & $375\pm 38$ & -- & \tablenotemark{a}  \\
MWA  & 158 & $340\pm 34$ & -- & \tablenotemark{a}   \\
MWA  & 166 & $291\pm 29$ & -- & \tablenotemark{a}   \\
MWA  & 174 & $315\pm 32$ & -- & \tablenotemark{a}  \\
MWA  & 181 & $248\pm 25$ & -- & \tablenotemark{a}  \\
MWA  & 189 & $353\pm 35$ & -- & \tablenotemark{a}  \\
MWA  & 197 & $284\pm 28$ & -- & \tablenotemark{a}  \\
MWA  & 204 & $335\pm 34$ & -- & \tablenotemark{a}   \\
MWA  & 212 & $312\pm 31$ & -- & \tablenotemark{a}   \\
MWA  & 216 & $269\pm 27$ & -- & \tablenotemark{a}  \\
MWA  & 220 & $287\pm 29$ & -- & \tablenotemark{a}  \\
MWA  & 227 & $310\pm 31$ & -- & \tablenotemark{a}  \\
GMRT & 617 & $200\pm 40$ & 4.0 & \tablenotemark{b}  \\
ASKAP& 943 & $140\pm 7$  & 5.5 & \tablenotemark{c}  \\
VLA  & 1400& $114\pm 4$  & 6.5 & \tablenotemark{e}  \\
VLA  & 1490& $104\pm 2$  & 7.1 & \tablenotemark{d}  \\
VLA  & 1570& $102\pm 5$  & 7.2 & \tablenotemark{c}  \\
VLA  & 4860& $45\pm 1$   & 14.6& \tablenotemark{d}  \\
VLA  & 6000& $38 \pm 2$  & 16.9& \tablenotemark{c}  \\
\enddata
\tablenotetext{a}{From the GLEAM source catalog \citep{Hurley-Walker+17}.} \tablenotetext{b}{\citet{irwin2013b}.} \tablenotetext{c}{From this work.} \tablenotetext{d}{\citet{heesen_2018b}.} \tablenotetext{e}{\citet{white1992}.}
\end{deluxetable*}

\begin{figure}[!htbp]
    \centering
    \includegraphics[width=0.98\columnwidth]{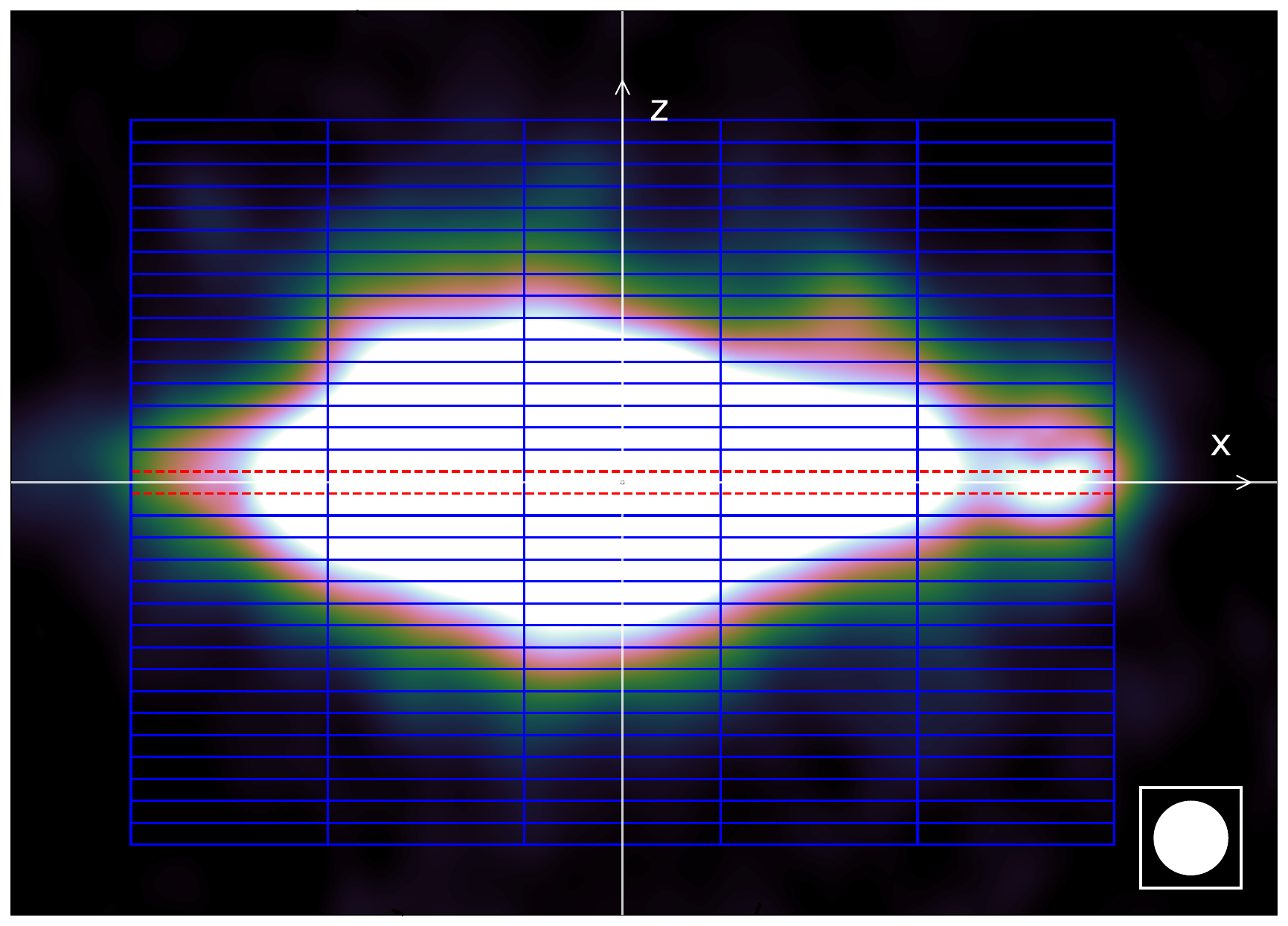}
    \caption{Rotated ASKAP \textbf{synchrotron emission intensity} image of NGC~3044. The definition of $x$ and $z$ axes is outlined. The vertical strips split into rectangles are also overlaid. Each rectangular box has a width of $43\arcsec$ (4.2~kpc) and a height of $5\farcs3$ (0.5~kpc).} 
\label{fig:cover}
\end{figure}

\subsection{Scale height of synchrotron emission}
\label{subsec:meth_bo_scal}

The scale heights for synchrotron emission at L and C-bands for NGC~3044 have already been determined~\citep[e.g.][]{krause2018,heesen_2018b}. The scale heights are derived only at 943~MHz from the ASKAP synchrotron intensity image, with the thermal emission subtraction detailed in Sect.~\ref{sec:therm_em}.

The measured intensity reflects both intrinsic vertical emission from the disk/halo and projected emission from the inclined disk for NGC~3044. Since we aim to measure the vertical profile, the line-of-sight contribution of the disk needs to be considered. We used the procedure by \citet{dumke1995} and \citet{heesen_2018b} to account for the broadening of the intrinsic vertical profile caused by the beamwidth ($\Theta_0$) of ASKAP and the contribution of the disk to the halo. 
A combined beamwidth $\Theta_c$ was introduced as $\Theta_c=\sqrt{\Theta_0^2+\Theta_d^2 \cos^2i}$, where $i$ is the inclination angle, $\Theta_d=R\cos(r/R\cdot\pi/2)$ is the equivalent beamwidth caused by the disk, $R$ is the radius of the disk, and $r$ is the distance from the center~\citep{muller2017}. The intrinsic Gaussian or exponential vertical profile was then convolved with the combined beamwidth to fit the observed profile to obtain the scale heights.

\begin{figure*}[!htbp]	
    \includegraphics[width=0.98\textwidth]{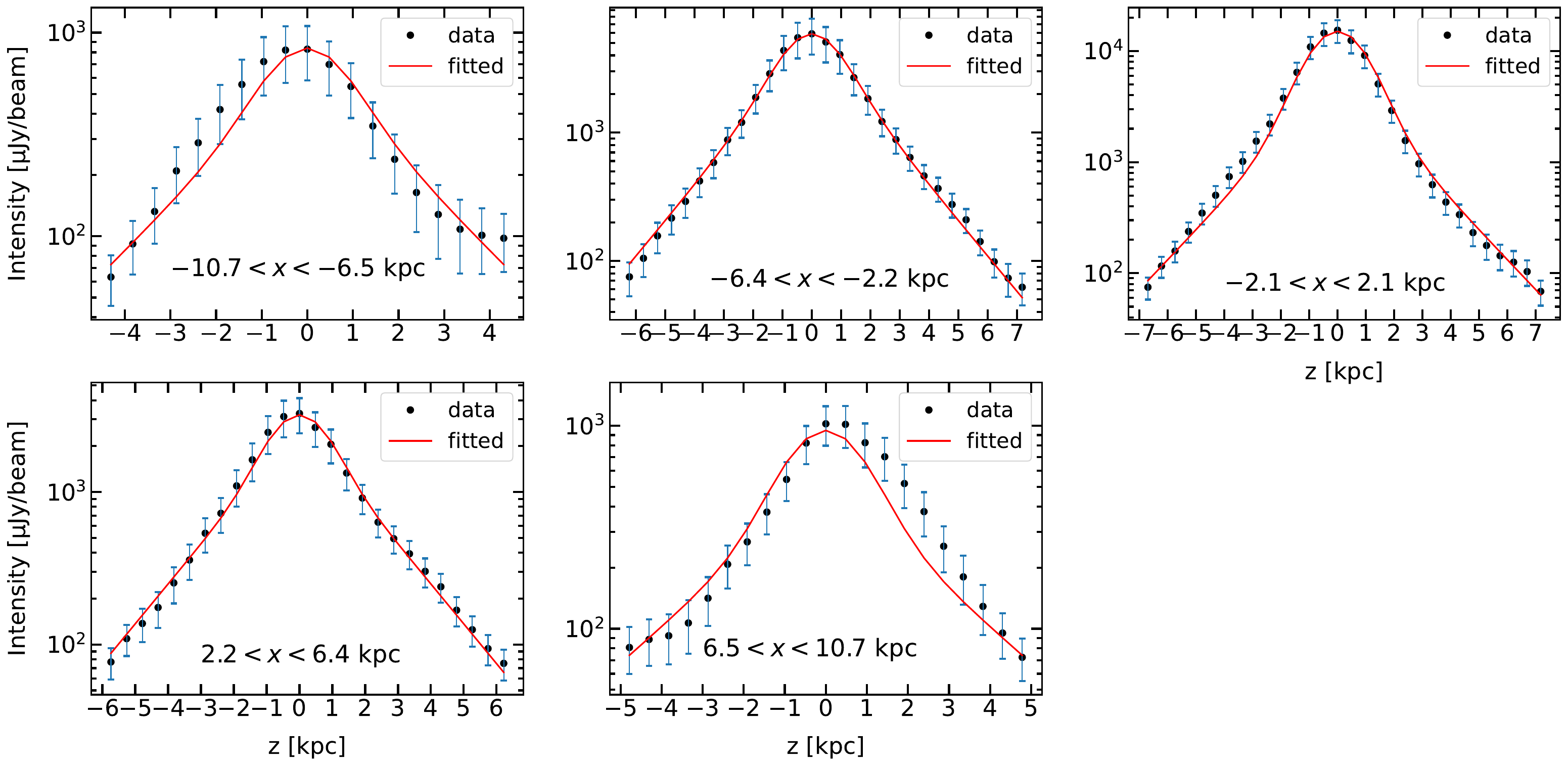}
    \caption{Intensity profiles of the strips outlined in Fig.~\ref{fig:cover}. The red curves show the two-component exponential fits, performed only on data points above $3\sigma$ ($\sigma=20\,\mathrm{\mu Jy\,beam^{-1}}$).} 
\label{fig:intensity-fit}
\end{figure*}

We rotated the ASKAP synchrotron emission intensity image of NGC~3044 clockwise by its position angle ($\rm PA=113\degr$) to align its major axis horizontally and defined the $x$ and $z$ axes parallel and perpendicular to the major axis, respectively, as shown in Fig.~\ref{fig:cover}. We then defined 5 strips along the $z$ axis, each with a width of $43\arcsec$ along the $x$ axis, corresponding to 2.7 times the beamwidth of the ASKAP image, ensuring sufficient independent values. Each strip was divided into rectangles of $5\farcs3$ height. Intensities were averaged within each rectangle and vertical profiles of intensity versus $z$ were derived for all 5 strips (Fig.~\ref{fig:intensity-fit}).

\begin{figure}[!htbp]	
    \centering
    \includegraphics[width=0.98\columnwidth]{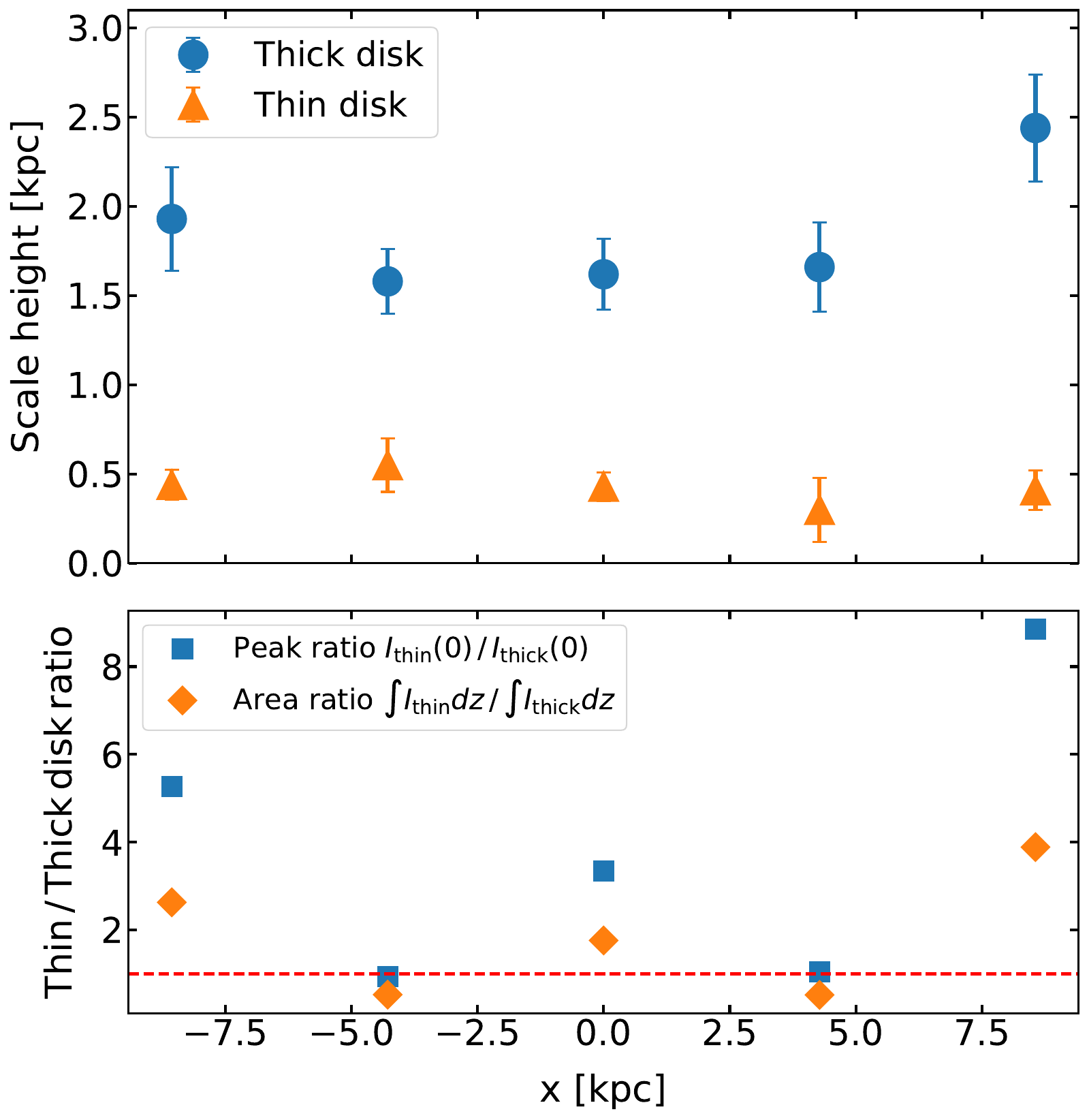}
    \caption{Top: Scale heights of the thin disk (blue circles) and thick disk (orange triangles) of NGC~3044 at 943~MHz. Bottom: Thin-to-thick disk intensity ratios as a function of position along the major axis, showing the mid-plane (peak) ratio, $I_{\rm thin}(0)/I_{\rm thick}(0)$, and the vertically integrated (area) ratio. The dashed line marks equal contributions from the two components.} 
\label{fig:scale_h}
\end{figure}

We found that an exponential function provides a better fit to the intensity profiles than a Gaussian function, and two exponential components, referred to as thin and thick disk components, are required for the fit. We used the reduced $\chi^2$ as a figure of merit to assess the fit. The scale heights of the thin and thick disks are nearly constant with $x$~(top panel in Fig.~\ref{fig:scale_h}). The corresponding averages are $0.43 \pm 0.13$ kpc and $1.91 \pm 0.26$ kpc, respectively. We quantify the relative contributions of the thin and thick disk components using the ratios of their peak intensities at $z=0$: $I_{\rm thin}(0)/I_{\rm thick}(0)$ and the ratios of their areas: $\int I_{\rm thin}{\rm d}z/I_{\rm thick}{\rm d}z$, shown in Fig.~\ref{fig:scale_h} (bottom panel). The thin disk dominates the mid-plane emission at all positions along the major axis. For the vertically integrated emission indicated by the area, the thin disk remains dominant in most regions. However, in two strips at $x\approx\pm5$~kpc, the thin disk contributes slightly less, with the thin-to-thick disk area ratio falling below unity, implying that the emission from the thin disk is not uniform across the disk and might be connected to galactic structures.

For NGC~3044, the thin and thick disk scale heights are $\simeq 0.1$ and $1.4$~kpc at 1.5~GHz, and $\simeq 0.1$ and $1.0$~kpc at 6~GHz, respectively~\citep{krause2018}. The thin-disk scale heights with large uncertainties (50–100\%) due to calibration errors and are further limited by the telescope angular resolution. The thick disk scale heights have uncertainties of 5–20\%, and the angular resolution is sufficient to separate this component, enabling reliable estimates. The thick disk (halo) is predominantly contributed by the synchrotron emission. For most edge-on galaxies, the synchrotron scale height is better described by a two-component exponential profile \citep{krause2018,heesen2025}, and the halo scale height decreases with increasing frequency, with averages of $1.4 \pm 0.7$~kpc at L-band and $1.1 \pm 0.3$~kpc at C-band \citep{krause2018}. At lower frequencies, the thick-disk scale height is larger because CREs have longer radiative loss times and thus propagate farther from the disk, producing a more extended halo.

 In this work, we assume that the synchrotron-emitting CREs are predominantly primary. Contributions from secondary electrons produced in hadronic interactions are expected to be small in normal star-forming disk galaxies such as NGC~3044, which lacks a compact starburst nucleus or AGN. Previous studies have shown that secondary CREs contribute only a minor fraction of the total synchrotron emission at GHz frequencies in such systems~\citep[e.g.][]{strong2007,lacki2010} and therefore do not affect our conclusions on CRE transport.

\section{Discussions}
\label{sec:discussions}
\subsection{Thermal emission}
\label{sec:therm_em}
Thermal emission in the radio band is dominated by free–free emission that can be estimated using the ``mixture method'' by combining the $\rm H\alpha$ and $\rm 24\,\mu m$ data. Star formation can be traced by these two types of radiation in both obscured and unobscured regions \citep{kennicutt2007}. The corrected H$\alpha$ flux $F_{\rm H\alpha}$ can be derived from the observed flux $F_{\rm H\alpha,\,obs}$ and the 24~$\mu$m intensity $I_{\rm 24\,\mu m}$ as~\citep{kennicutt2009}:
\begin{equation}
    F_{\rm H\alpha} = F_{\rm H\alpha,obs} + 0.042 \cdot \nu_{24\,\rm \mu m}\, I_{24\,\rm{\mu m}}.
    \label{Halpha+24microns}
\end{equation}

We used the H$\alpha$ data published by \citet{vargas2019}. There are no $\rm 24\,\mu m$ observations available for NGC~3044, and we used the Wide-field Infrared Survey Explorer (WISE) W4 ($\rm 22\,\mu m$) image as a substitute~\citep{wise}. There is a tight linear relation for the emission between the two wavelengths, and a factor of 1.03 was multiplied to obtain the extrapolation from 22~$\mu$m~\citep[e.g.][]{wiegert2015}.

The thermal contribution to the radio continuum emission at a given frequency $\nu$ can then be calculated using~\citep{deeg1997}
\begin{align}
\frac{S_{\mathrm{th}}(\nu)}{\mathrm{erg\,cm^{-2}\,s^{-1}\,Hz^{-1}}}
&= 1.14 \times 10^{-14}
   \left( \frac{\nu}{\mathrm{GHz}} \right)^{-0.1} \nonumber \\
&\quad \times \left( \frac{T_e}{10^4\,\mathrm{K}} \right)^{0.34}
   \left( \frac{F_{\mathrm{H}\alpha}}{\mathrm{erg\,cm^{-2}\,s^{-1}}} \right).
\label{radioequation}
\end{align}
The electron temperature $T_e$ was assumed to be 10,\,000~K.

\begin{figure*}[!htbp]	
    \centering
    \includegraphics[width=\linewidth]{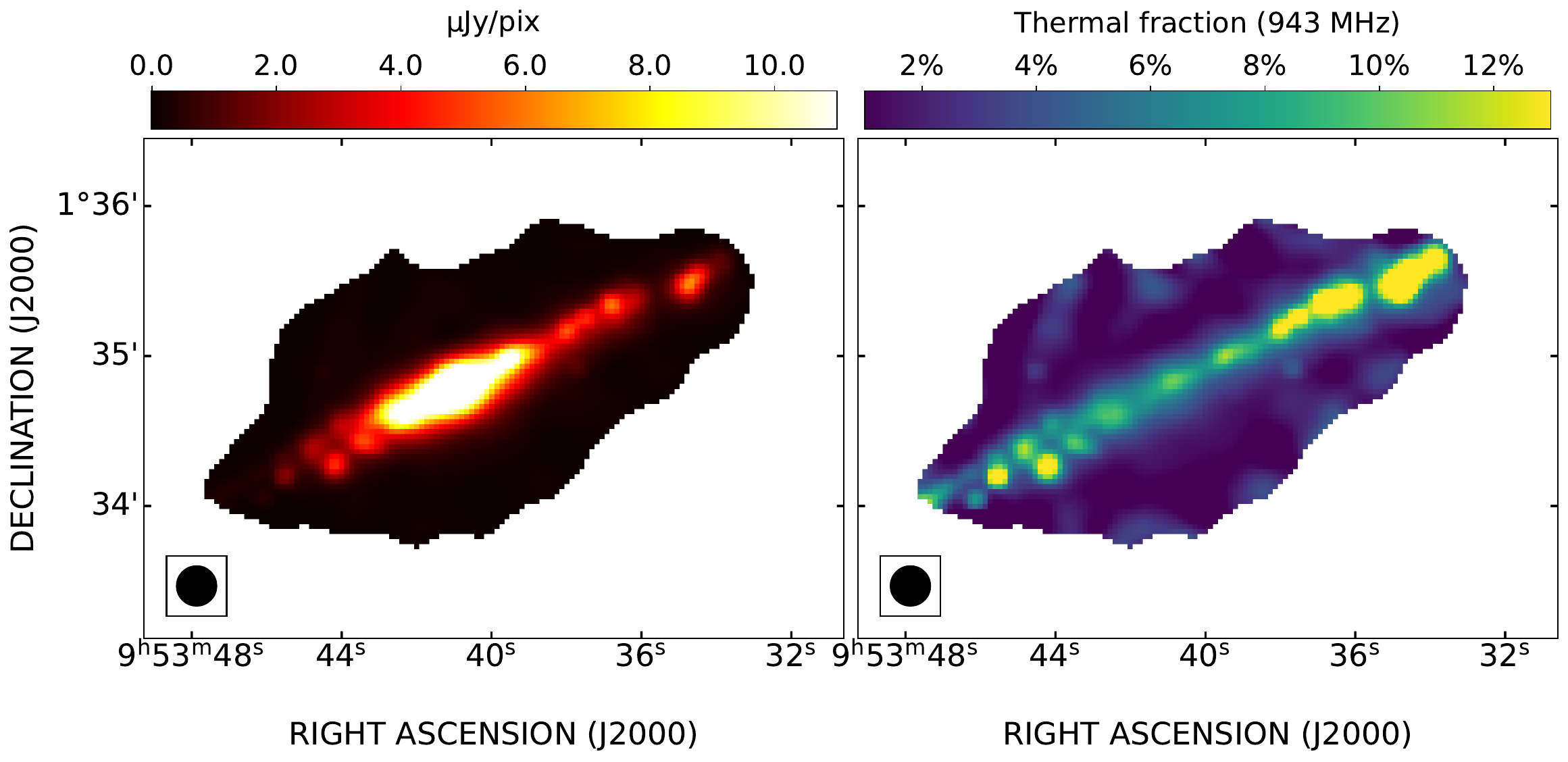}
    \caption{Thermal emission distribution at 943 MHz (left) and the corresponding thermal fraction (right).} 
\label{fig:Thermal_emission}
\end{figure*}

We derived the thermal emission distribution at 943~MHz and the corresponding thermal fraction, as shown in Fig.~\ref{fig:Thermal_emission}. There is a prominent thermal emission ridge along the major axis of the galaxy, which peaks at the center. The thermal emission distributions at L and C-bands exhibit similar structures~\citep{vargas2018}. 

Using the integrated flux density, we estimated the contribution of the thermal fraction at different frequencies, and assuming that the MWA is entirely from synchrotron emission. The results are presented in Table~\ref{tab:int_flux}. Due to its low star formation rate, the thermal fraction in this galaxy is relatively low. Nevertheless, since our focus is on the analysis of non-thermal synchrotron emission, we subtract the contribution of the thermal component from the total intensity maps in the subsequent analysis.

\subsection{Spectral index}
\label{subsec:spex}
\subsubsection{Integrated flux density spectrum}
We show the integrated flux density versus frequency from the data in Table~\ref{tab:int_flux}, excluding the thermal contribution estimated in Sect.~\ref{sec:therm_em}. 

\begin{figure}[!htbp]
    \centering
    \includegraphics[width=0.98\columnwidth]{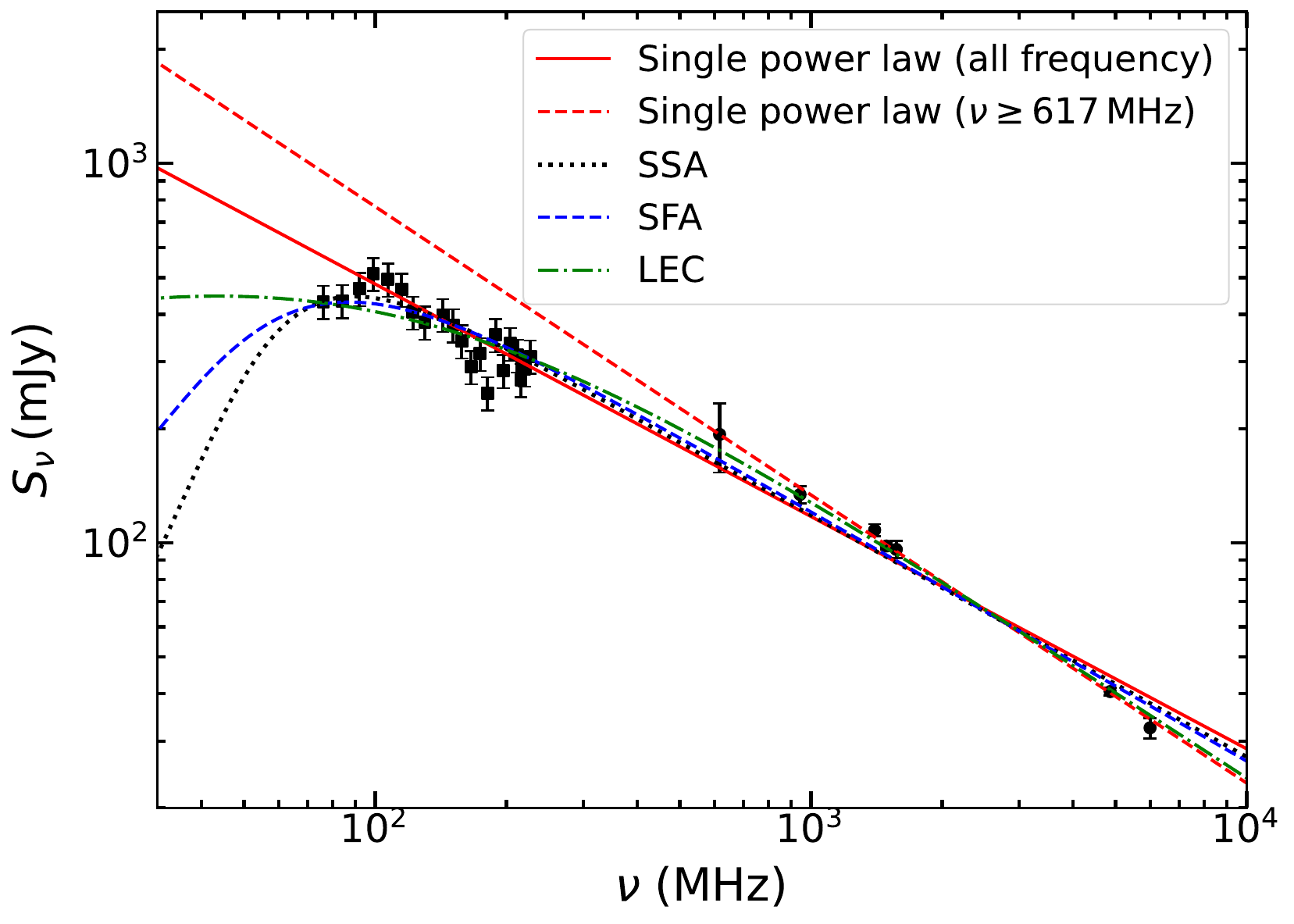}
    \caption{Thermal component subtracted integrated flux density spectrum fitted with models of a single power law (red solid line); the red dashed line represents a fit using only the black filled circles; synchrotron emission with self-absorption (SSA, black dotted line), synchrotron emission with free-free absorption (SFA, blue dashed line), and synchrotron emission with low energy cutoff of electrons (LEC, green dot-dashed line).
} 
\label{fig:spectral_index_fit}
\end{figure}

We fit all the data to a single power law $S_\nu\propto\nu^\alpha$ with $\alpha$ being the spectral index (Fig.~\ref{fig:spectral_index_fit}, red solid line). We obtained a spectral index of $-0.61 \pm 0.01$. The spectral index shows a clear flattening at low frequencies ($<300$~MHz). As a test, we refitted the spectrum after excluding the MWA/GLEAM data, obtaining $\alpha=-0.76\pm0.02$. This value is fully consistent with the $\alpha=-0.75$ reported by~\cite{heesen_2018b} between 1.49 and 4.86~GHz, and is also very similar to the $-0.8$ reported by~\cite{irwin2024}.

The spectrum appears to be inverted at lower frequency ($<100$~MHz), which could be modeled by synchrotron emission with self-absorption~\citep[SSA,][]{tingay2003}, synchrotron emission with free-free absorption~\citep[SFA,][]{mcdonald2002,kapinska2017}, and synchrotron emission with low-energy cutoff of electrons~\citep[LEC,][]{granot2002}. We tried to fit the spectrum with these three models, shown in Fig.~\ref{fig:spectral_index_fit}. 

For the SSA model, the spectrum can be represented as 
\begin{equation}
S_{\nu} = S_{\tau=1}\left(\frac{\nu}{\nu_{\tau=1}}\right)^\alpha
\left(\frac{1 - e^{-\tau(\nu)}}{\tau(\nu)}\right),
\label{eq:ssa}
\end{equation}
where the optical depth depends on the frequency and the spectral index as $\tau(\nu) = \left( \nu/ \nu_{\tau=1} \right)^{\alpha-2.5}$. The fitting yielded $S_{\tau=1}=651\pm26\,\rm mJy$, $\alpha=-0.64\pm0.01$, and $\nu_{\tau=1}=70\pm6$~MHz.

For the SFA model, the spectrum is written by
\begin{equation}
S_{\nu} = S_0\left(\frac{\nu}{\nu_0}\right)^\alpha
\left(\frac{1-e^{-\tau_{\rm ff}(\nu)}}{\tau_{\rm ff}(\nu)}\right),
\label{eq:ffa}
\end{equation}
where the optical depth depends on the frequency as $\tau_{\mathrm{ff}}(\nu) = \left( \nu/ \nu_{\tau=1} \right)^{-2.1}$, and $\nu_0=1$~GHz is a reference frequency. The fitted parameters are $S_0 = 121 \pm 2$~mJy, $\alpha = -0.66 \pm 0.01$, and $\nu_{\tau=1}=74\pm8$~MHz.

For the LEC model, the spectrum can be described as 
\begin{equation}
\label{eq:lec}
S_{\nu}
= A\left(\frac{\nu}{\nu_{\min}}\right)^{1/3}
\left[1+\left(\frac{\nu}{\nu_{\min}}\right)^{\frac{1/3+\alpha}{s}}\right]^{-s},
\end{equation}
where $A=2^s\,S_{\nu_{\rm min}}$ is a normalization, $S_{\nu_{\rm min}}$ is the flux density at $\nu_{\rm min}$, and $\nu_{\min}$ is the characteristic frequency associated with the cutoff Lorentz factor $\gamma_{\min}$ of electrons, and $s$ controls the smoothness of the transition. The best fit resulted in $\alpha=-0.76\pm0.02$ and $\nu_{\rm min}=93\pm17$\,MHz (for $s=1$), leading to $S_{\rm min}=414$~mJy.

All three models reproduce the data above $\sim100$\,MHz but diverge below $\sim70$\,MHz (Fig.~\ref{fig:spectral_index_fit}). The fitted optically thin spectral indices are similar ($\alpha\simeq-0.61$ to $-0.80$), confirming a stable non-thermal slope once the thermal component is removed. The key discriminator is the turnover mechanism: SSA is typically dominant in compact, high-brightness sources~\citep[e.g. AGN cores, radio supernovae, or compact starburst nuclei;][]{slysh1990,chevalier1998,kapinska2017}. However, NGC~3044 shows no evidence for an AGN or a compact starburst nucleus. An LEC with $\nu_{\min}\simeq90$\,MHz corresponds to $\gamma_{\min}\simeq1.5\times10^{3}$ ($E_{\min}\simeq0.75$\,GeV) for $B\simeq10\,\mu$G; nevertheless, a uniform, galaxy-wide cutoff is unlikely because spatial variations in $B$ and line-of-sight mixing would smear any sharp turnover, and sub-GeV electrons are generically produced. We therefore regard SFA as the most plausible origin of the low-frequency turnover, while allowing for minor contributions from localized SSA or gentle low-energy softening at the tens-of-MHz level. New data below $\sim$70\,MHz would decisively distinguish these scenarios.

\subsubsection{Spectral index map}

\begin{figure*}[!htbp]
    \centering
    \includegraphics[width=\linewidth]{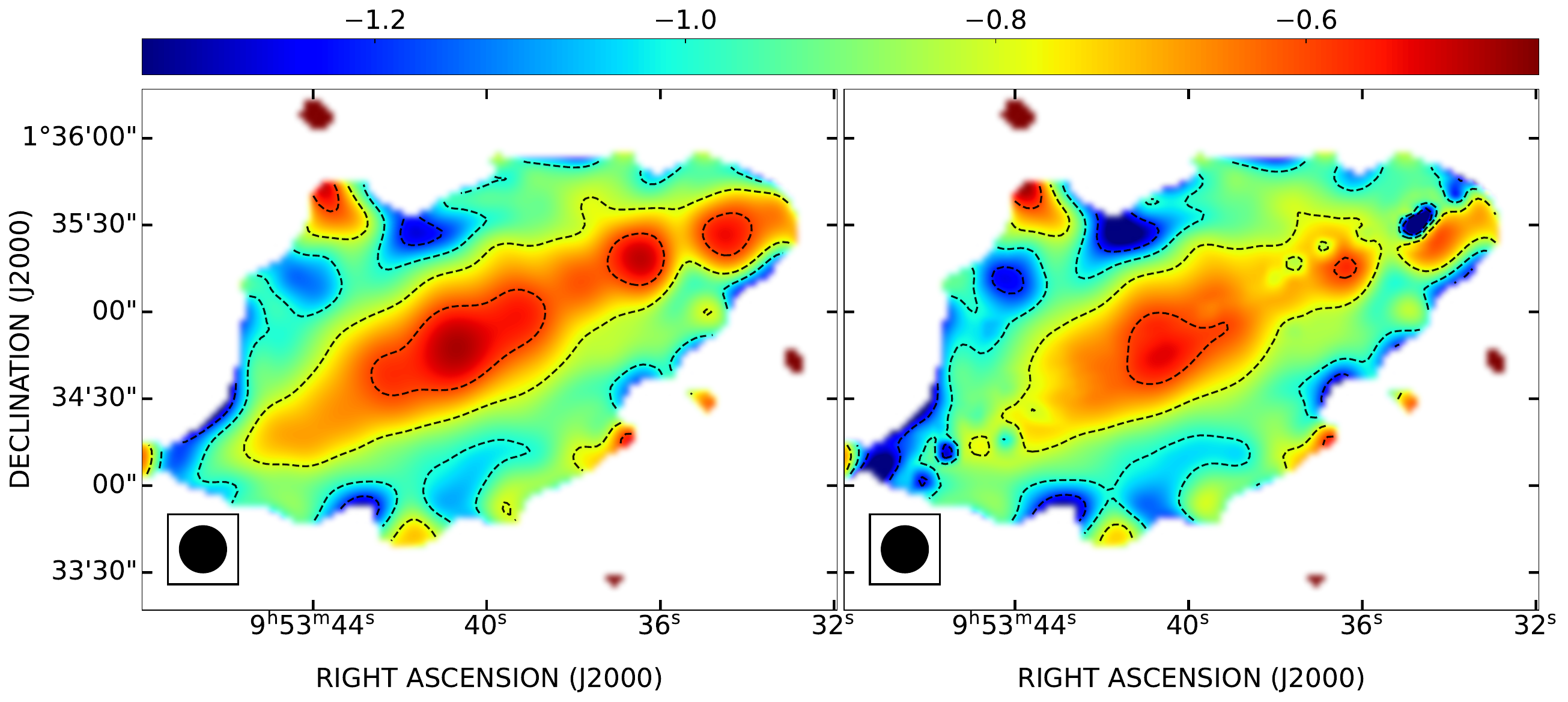}
    \caption{Spectral index map for total intensity (left) and synchrotron emission (right). The black circles in the lower left corner indicate the beam size of $16\arcsec$. The contour levels are $-$1.2, $-$1.0, $-$0.8, and $-$0.6. Regions with $I<3\sigma$ were excluded.} 
\label{fig:spex_distrub}
\end{figure*}

We derived a pixel-based spectral index map for total intensity~(Fig.~\ref{fig:spex_distrub}, left) using the ASKAP image at 943~MHz and the VLA images at 1.57~GHz and 6~GHz, assuming a single power-law model. We subtracted thermal emission pixel by pixel based on the estimate in Sect.~\ref{sec:therm_em} and derived the spectral index map for synchrotron emission (Fig.~\ref{fig:spex_distrub}, right panel). The spectral index was derived only when all the intensities at the three frequencies are larger than three times the rms noise.

The spectral index of the synchrotron emission near the galactic disk is steeper than that of the total intensity. This is expected since the total intensity also contains thermal emission that has a flatter spectrum. The spectral index of synchrotron emission gradually steepens from the disk to the halo, indicating that CREs continuously lose energy as they propagate outward.

\subsection{Magnetic field strength}
\label{sec:equi_B}

\begin{figure}[!htbp]	
    \centering
    \includegraphics[width=0.98\columnwidth]{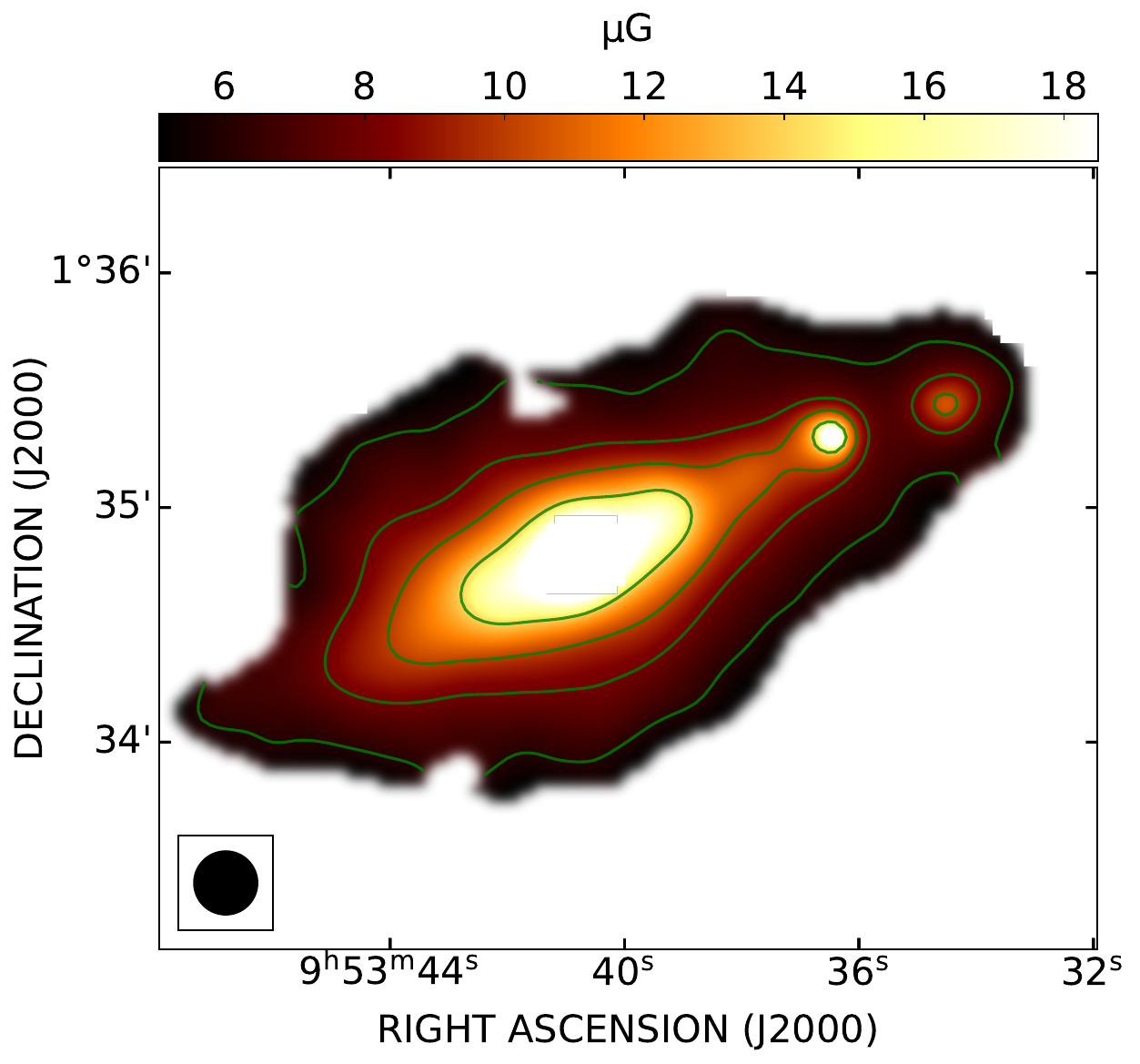}
    \caption{Map of the total magnetic field strength derived from energy equipartition. Contours are at 6, 8, 10, 14~$\rm \mu G$.} 
\label{fig:B_distri}
\end{figure}

We estimated the magnetic field strength under the assumption of energy equipartition between CRs and the magnetic field \citep{beck2005}:
\begin{equation}
	B_{\mathrm{eq}} = \left( \frac{ 4\pi (2\alpha + 1) (K_0 + 1) I_\nu  E_p^{1-2\alpha} \left(\frac{\nu}{2c_1}\right)^{\alpha} }{ (2\alpha - 1) c_2(\alpha) L \cdot c_4(i) } \right)^{\frac{1}{3 + \alpha}}, \label{eqn:B_eq}
\end{equation}
where $c_1$ is a constant, $c_2$ depends on the spectral index $\alpha$, $E_p$ is the spectral break energies of protons ($E_p=938.28\,\rm MeV=1.5\times 10^{-3}\,\rm erg$), $I_{\nu}$ is the synchrotron intensity in $\rm erg\,s^{-1}\,cm^{-2}\,Hz^{-1}\,sr^{-1}$, $L$ is the effective path length through the source, $K_0$ is the ratio of the proton energy to the electron energy, and $c_4=(2/3)^{(1+\alpha)/2}$ assuming an isotropic magnetic field. For an isotropic turbulent field, $c_4 \approx 0.7$ for $\alpha = -0.8$, while $c_4 = 1$ represents the limiting case of a perfectly regular field aligned in the plane of the sky. Even in this extreme and physically unlikely case, the resulting equipartition field strength differs by only $\sim9\%$, which is an upper limit due to the uncertainty of magnetic field geometry.

To estimate the average magnetic field strength for the entire galaxy, we adopted the spectral index $\alpha = -0.61$ obtained above and the ratio $K_0 = 100$. For path length, we used the mean full width at half power (FWHP) of the fitted vertical synchrotron profiles, $L\approx4$~kpc. With these assumptions, the inferred magnetic field strength is 8.8~$\mu$G, consistent with values reported for many spiral galaxies~\citep{beck2000,heesen2022}.

To derive the magnetic field strength map, we used the ASKAP at 943~MHz total intensity image after subtracting the thermal emission and the spectral index map of the synchrotron emission~(Fig.~\ref{fig:spex_distrub}, right panel) and the same $K_0$ and $L$ as above. To avoid overestimating the magnetic field strength, we only used areas with the spectral index in the range $-1.2 < \alpha < -0.5$, following the discussion by~\citet{heesen2022}. The resulting distribution of the magnetic field strength is shown in Fig.~\ref{fig:B_distri}. It can be clearly seen that the magnetic field strength declines from 16~$\mu$G at the galactic midplane to 6~$\mu$G at the halo. 

\subsection{Advection and diffusion: fitting NGC~3044 observations}
\label{subsec:spin_fit}

The transport of CREs from the disk to the halo operates in two modes, advection and diffusion. The advection-dominated transport generates a more extensive radio halo with less cooling for CREs compared to that of the diffusion-dominated transport. These modes can be differentiated using profiles of synchrotron intensity and spectral index~\citep{heesen2016, heesen_2018b, heesen2021}.

\begin{figure*}[!htbp]
    \centering
    \includegraphics[width=0.47\textwidth]{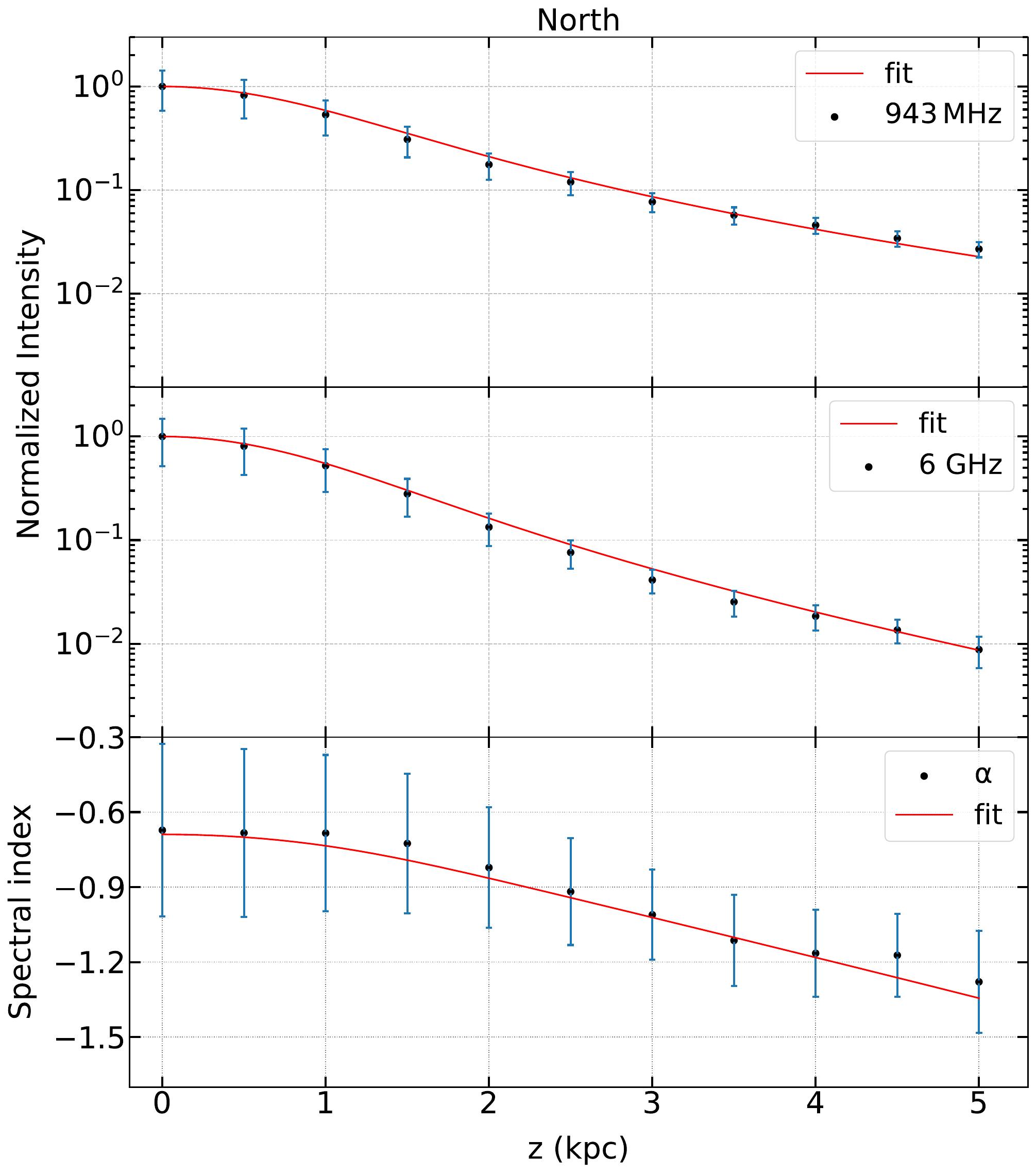}
    \includegraphics[width=0.47\textwidth]{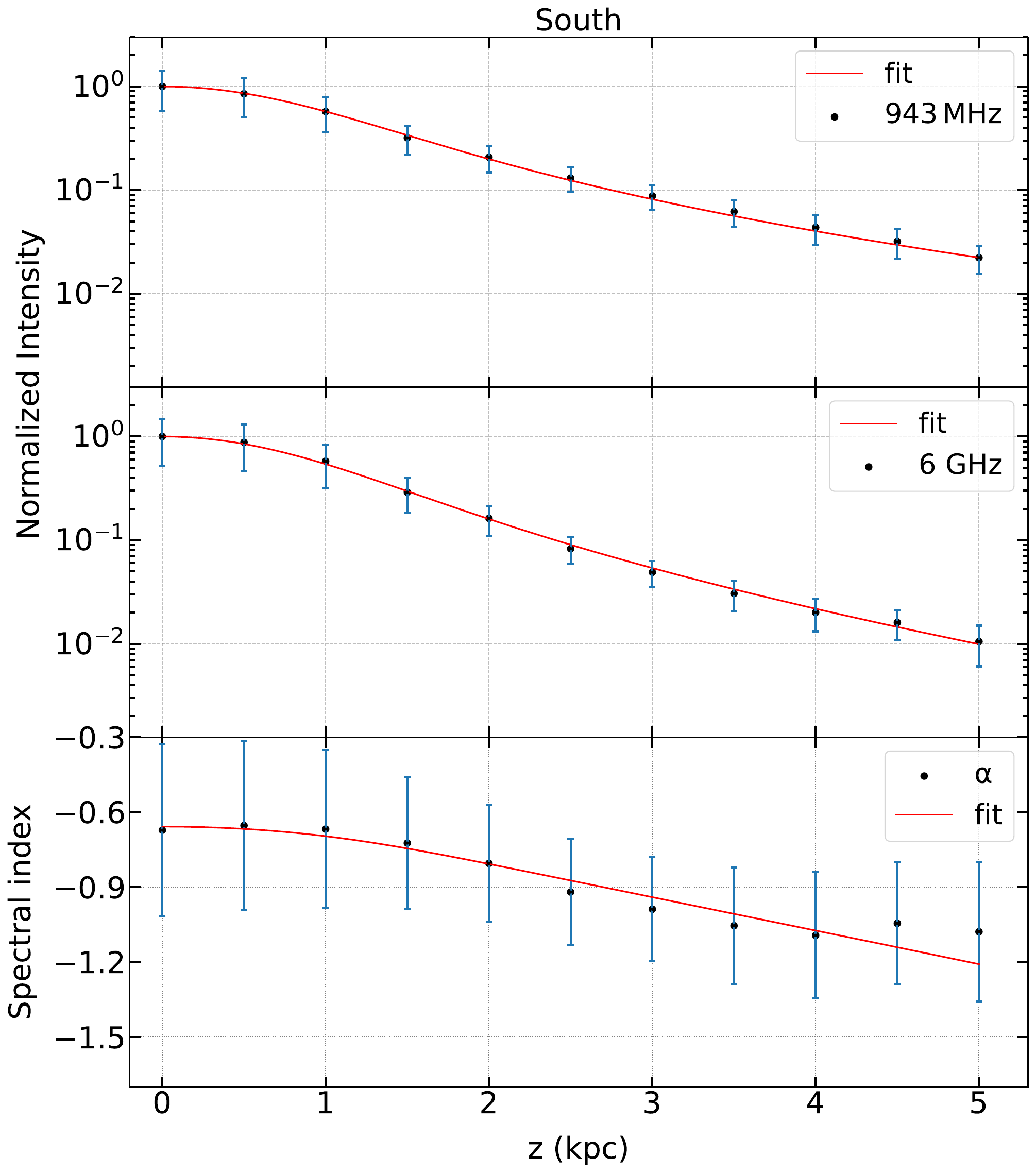}
    \caption{Profiles of synchrotron intensity and spectral index. Red lines indicate the fitting with the advection mode of CRE transportation.}
    \label{fig:profiles_fit}
\end{figure*}

We derived the synchrotron intensities at 943~MHz and 6~GHz after removing the thermal contribution, and then derived the corresponding spectral index for the northern and southern hemispheres when rotated to align the major axis horizontally. Vertical profiles (Fig.~\ref{fig:profiles_fit}) were created and normalized by the values at $z=0$. We used \textrm{{\small SPINNAKER}}\footnote{\url{https://github.com/vheesen/Spinnaker}} to solve the 1D advection and diffusion equations along $z$ and fit the solutions to these profiles. The magnetic field was modeled with a two-component exponential function:
\begin{equation}
   B(z)=B_1\,\mathrm{exp}\left(-\frac{|z|}{h_{\rm B1}}\right)+B_2\,\mathrm{exp}\left(-\frac{|z|}{h_{\rm B2}}\right).
\label{eq:exp_B}
\end{equation}
We obtained the initial values of $B_1=6$~$\mu$G and $B_2=7$~$\mu$G by fitting the equipartition magnetic field (Fig.~\ref{fig:B_distri}) and fixed these values when using SPINNAKER. 

We found that advection better reproduced the observed profiles of synchrotron intensity and spectral index for the north and south hemispheres, as shown in Fig.~\ref{fig:profiles_fit}. The preferred velocity profile follows a power law as
\begin{equation}
  v=v_0\left[ 1+\left(\frac{z}{h_v}\right)^\beta\right].
\label{eq:adv_v} 
\end{equation}
The fitted parameters of $h_{B1}$, $h_{B2}$, $v_0$, $\beta$, and $h_v$ are listed in Table~\ref{table:spin_fit_para}. The parameters for the two hemispheres are very similar. 

\begin{table}[!htbp]
\caption{Best-fitting parameters for the advection of the CREs.}
\centering
\label{table:spin_fit_para}
\begin{tabular}{lcc}
\hline
             & North         & South \\      
\hline
$h_{B1}$ (kpc)                       & $2.0\pm0.3$  & $2.0\pm0.3$\\
$h_{B2}$ (kpc)                       & $18\pm1$     & $18\pm1$\\
$\gamma_{\rm inj}$                   & $2.30\pm0.04$& $2.25\pm0.03$ \\
$v_0$ ($\rm km\,s^{-1}$)             & $105\pm10$   & $110\pm20$  \\
$h_v$ (kpc)                          & $1.4\pm0.2$  & $1.2\pm0.2$ \\ 
$\beta$                              & $1.23\pm0.05$  & $1.20\pm0.05$ \\
\hline
\end{tabular}
\end{table}

\begin{figure}[!htbp]
    \centering
    \includegraphics[width=0.98\columnwidth]{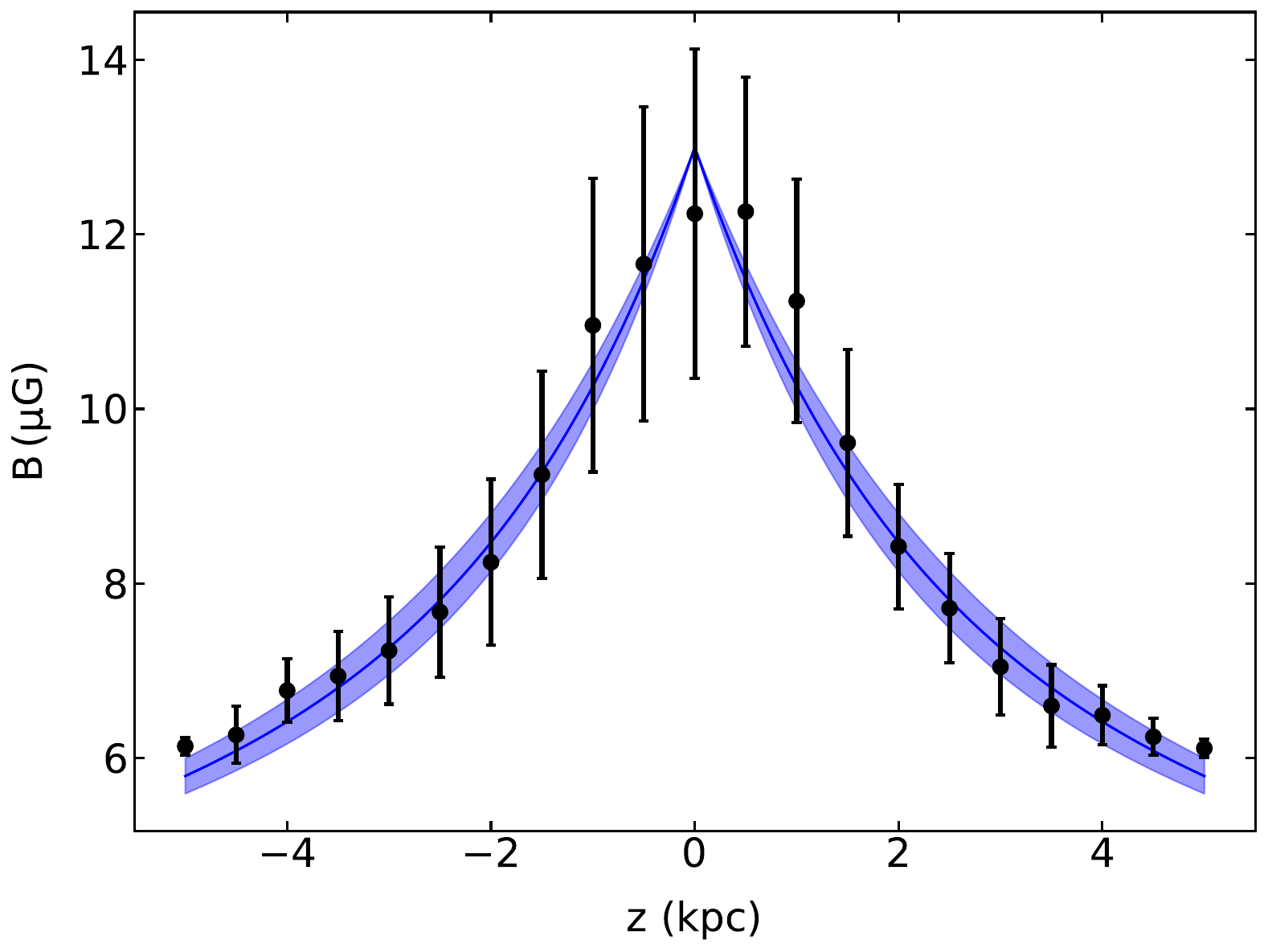}
    \caption{Equipartition magnetic field strength (dot) in comparison with that from SPINNAKER fitting (blue shaded area for $1\sigma$ error).} 
\label{fig:B_model}
\end{figure}

For consistency check, we show the profiles of the magnetic field estimated with energy equipartition and derived from SPINNAKER fitting in Fig.~\ref{fig:B_model}. As can be seen, these profiles agree well. This indicates that CREs have not experienced significant cooling and thus can trace the magnetic field properly, which is consistent with the propagation of CREs in advection mode.

\subsection{Galactic wind}
CREs are advected outward, indicating the existence of a galactic wind. If CRs provide the primary driving source of the galactic wind, the CREs will eventually escape the gravitational potential. 

\begin{figure}[!htbp]
    \centering
    \includegraphics[width=0.98\columnwidth]{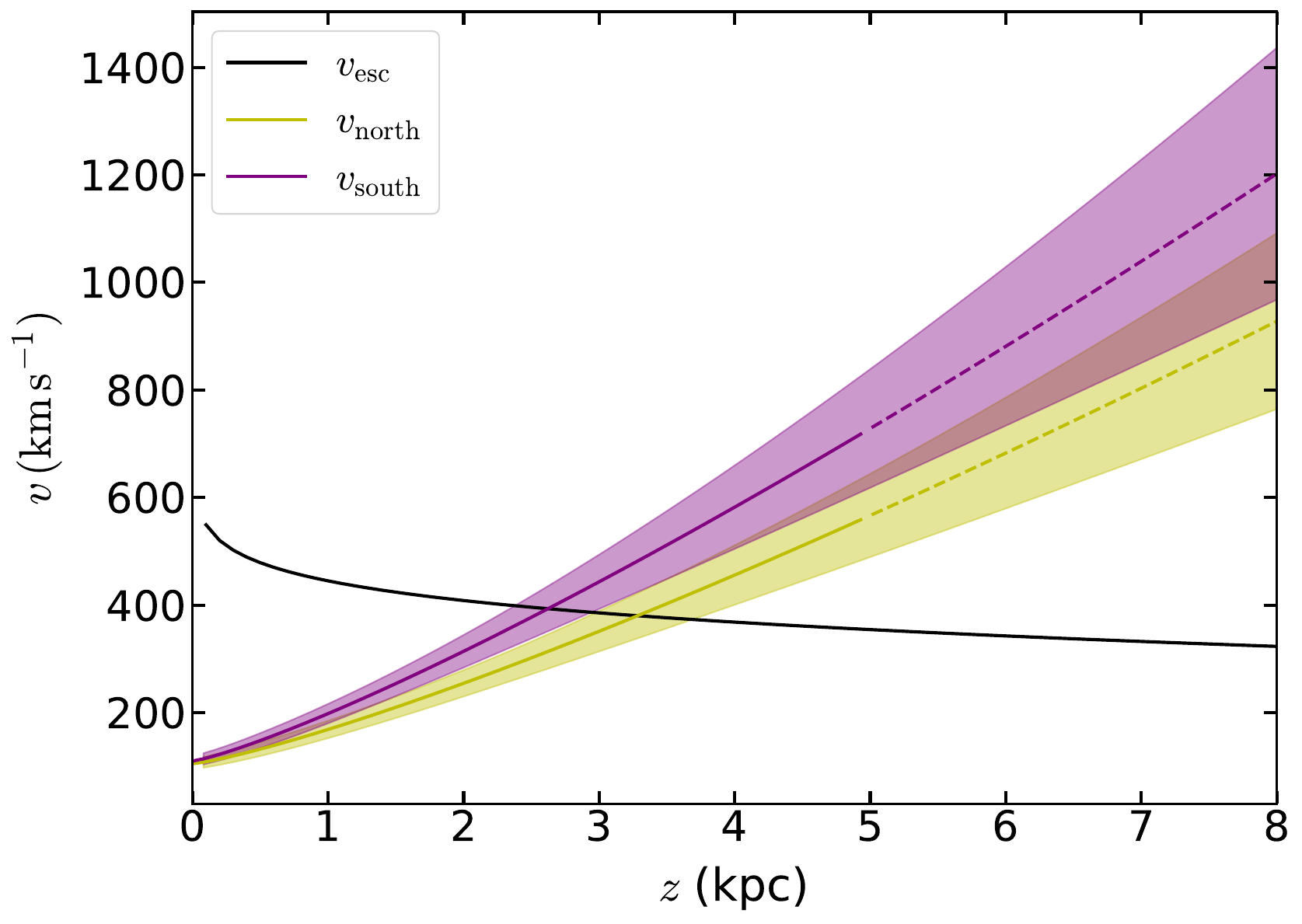}
    \caption{Vertical profiles of the escape velocity and the advection velocity for the north (yellow) and south (purple) hemispheres of NGC~3044. The shaded areas indicate the $1\sigma$ uncertainties. Dashed segments of the velocity curves denote the extrapolated portions of the profiles beyond the range directly constrained by the radio data.}
\label{fig:gal_wind}
\end{figure}

Following \cite{veilleux2005} and \cite{stein2022}, the escape velocity $v_{\rm esc}$ can be written as
\begin{equation}
  v_{\rm esc}=\sqrt{2}\,v_{\rm rot}\sqrt{1+\mathrm{ln}\left(\frac{R_{\rm max}}{r}\right)},
\label{eq:esc_speed}
\end{equation}
where $v_{\rm rot}=153\,\mathrm{km\,s^{-1}}$ is the rotational velocity of NGC~3044~\citep{li2016}, $r$ is the spherical radius, $R_{\rm max}$ is the truncated iso-thermal dark matter halo radius with an assumed value of 30~kpc. The vertical profile of $v_{\rm esc}$ versus $z$ is shown in Fig.~\ref{fig:gal_wind}.

Vertical profiles of advection velocities for the north and south hemispheres of NGC~3044 derived from SPINNAKER fitting~(Table~\ref{table:spin_fit_para}) are also shown in Fig.~\ref{fig:gal_wind}. It can be seen that the galactic wind is launched from the disk ($z=0$) with a small initial velocity and accelerated into the halo, reaching the escape speed of $\sim400$~km~s$^{-1}$ at a height of about 3~kpc. Beyond this height, the wind exceeds the gravitational potential and escapes the galaxy.

\begin{figure}[!htbp]
    \centering
    \includegraphics[width=0.98\columnwidth]{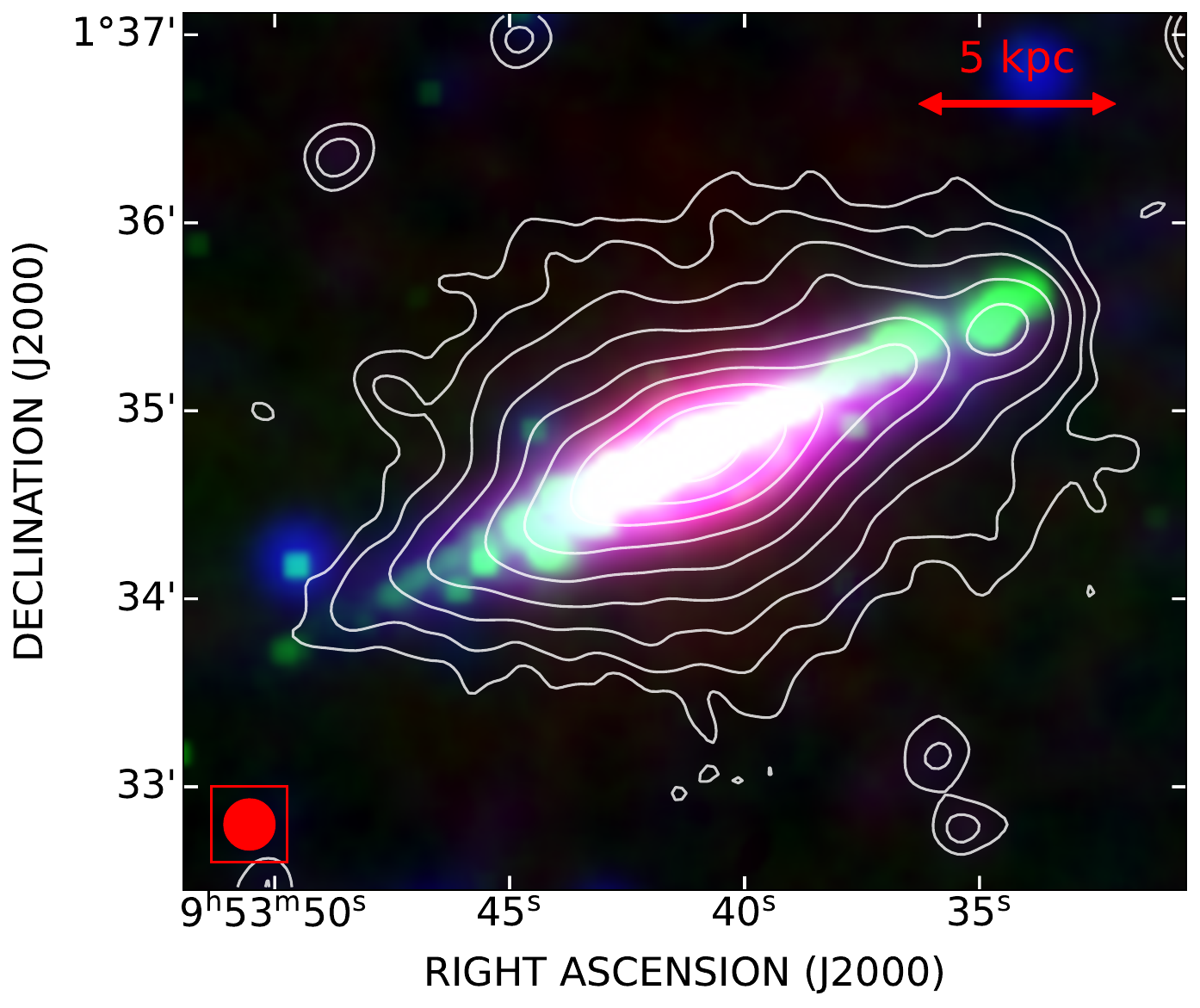}
    \caption{Three-color image of NGC 3044 with W1 (blue), $\rm H\alpha$ (green), and W3 (red) bands, overlaid with contours for radio total intensity.} 
\label{fig:w1_w3_color_map}
\end{figure}

We present a three-color image in Fig.~\ref{fig:w1_w3_color_map}, combining W1 ($3.4\,\rm \mu m$), W3 ($12\,\rm \mu m$), and H$\alpha$, with all bands convolved to a common $16\arcsec$ resolution. W3 (red) traces PAH-rich, warm dust heated by recent star formation and is less affected by extinction, while H$\alpha$ (green) traces unobscured ionized gas from very young massive stars. It is clearly seen that in NGC~3044, the redder center indicates dust-rich, partially obscured star formation in the nucleus/midplane, whereas H$\alpha$ is more prominent in the midplane and on either side of the central region. The pressure from the starburst drives the materials out of the disk, forming a ``superwind''. Once the wind enters the halo, the dominant driving agent changes from thermal pressure to CR pressure. As a result, at large vertical heights and without ongoing star formation, CR pressure can maintain the acceleration of the wind as seen in Fig.~\ref{fig:gal_wind}. 
 
The large propagation velocity of CREs has also been found for other galaxies with radio continuum observations, such as NGC~4666~\citep{heesen_2018a,stein_2019a}, NGC~5775~\citep{heald2022}, and NGC~253~(Wang et al. 2025, A\&A, submitted), all reaching the range of 400-700~km~s$^{-1}$. In all of these cases, it has been shown that the CR pressure plays a major role in lifting and accelerating the gas.

\subsection{A possible supperbubble}
\label{sec:supperbubble}

\begin{figure}[!htbp]
    \centering
    \includegraphics[width=0.98\columnwidth]{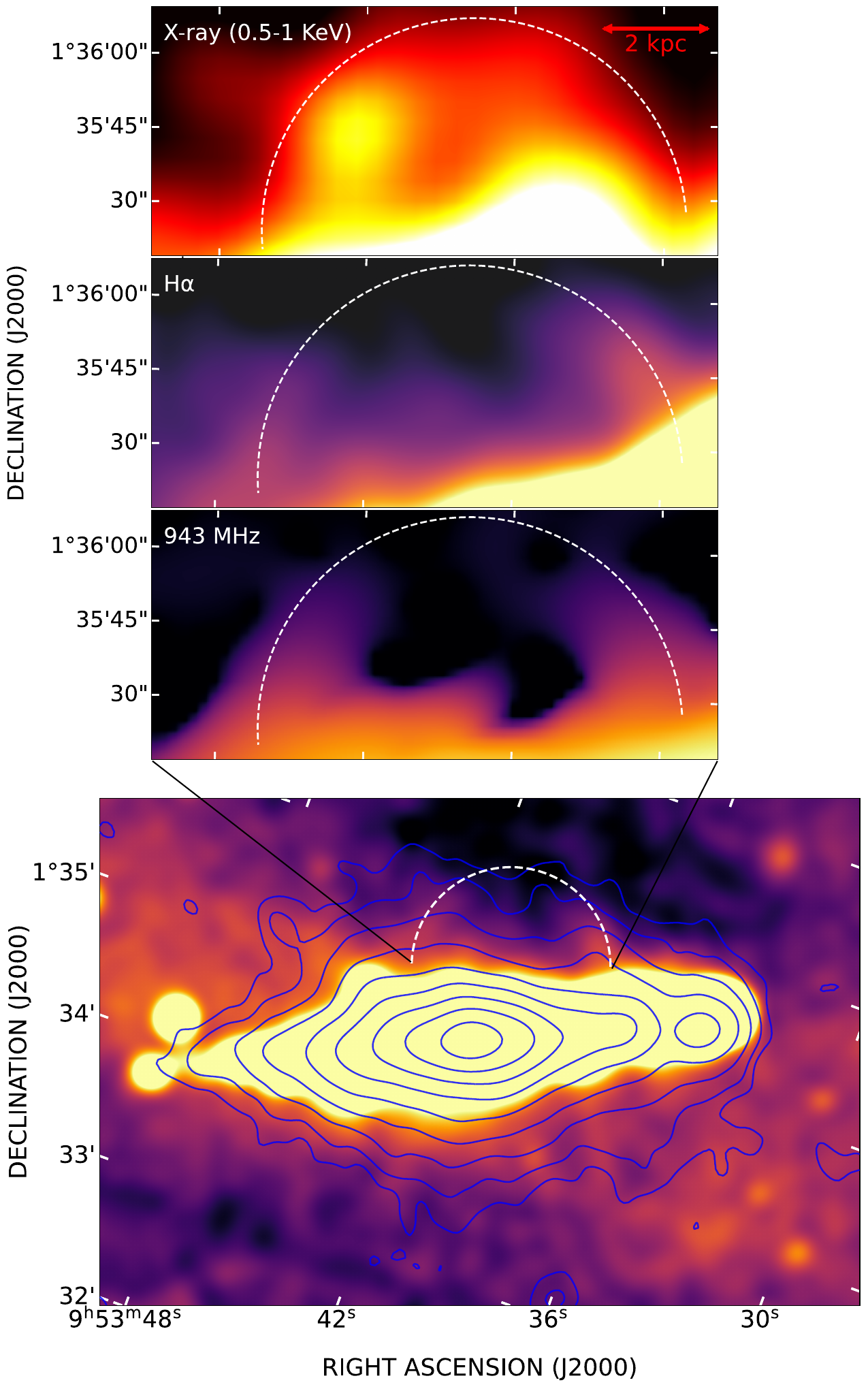}
    \caption{H$\alpha$ image overlaid with contours for total intensity at 943~MHz (bottom panel). The white dashed line marks the possible superbubble. X-ray emission from XMM-Newton (0.5-1~KeV), H$\alpha$, and 943~MHz total intensity for the area around the superbubble are shown in the zoom-in panels.} 
\label{fig:super_bubble}
\end{figure}

There appears to be a concave in the north of the center in the ASKAP total intensity image~(Fig.~\ref{fig:total_intensity}, top left panel), which resembles a bubble. We zoom in this area and show X-ray, H$\alpha$, and total intensity at 943~MHz in Fig.~\ref{fig:super_bubble}. For the zoom-in radio image, we used the multiscale decomposition method by~\citet{li2022} to filter out small-scale features. A partial shell structure with a breakout on top can be identified in both radio and H$\alpha$ images, which has a good correspondence between the two wavelengths. The X-ray emission fills the interior. The shell structure has a radius of $R\simeq3$~kpc and its center is at a height of about 2.5~kpc from the galactic plane and is offset by $\sim 3$~kpc from the galactic center along the major axis. We refer to this structure as a supperbubble in the following discussions.

The location of the superbubble is not symmetric about the galactic midplane. Yet producing a structure of this scale requires a large, concentrated energy input from massive stars or supernovae, which is unlikely to occur at a height of $\geq2.5$~kpc above the disk. A plausible explanation is that the bubble, while expanding outward from the disk, has been displaced upward by buoyancy in the surrounding gas.

To maintain the superbubble, the internal and external pressures need to be balanced. The external pressure $P_e$ consists of thermal pressure $P_{\rm th}=nkT$, turbulent pressure $P_{\rm turb}=\rho\sigma_{\rm turb}^2$, magnetic pressure $P_{\rm B}=B^2/8\pi$, and CR pressure $P_{\rm CR}=P_{\rm B}$ assuming energy equipartition. Here, $n$ is the number density, $k$ is the Boltzmann constant, $T$ is the temperature, $\rho=\mu_{\rm H}m_{\rm H}n$ is the mass density with $\mu_{\rm H}=1.4$ and $m_{\rm H}$ being the hydrogen atom mass, $\sigma_{\rm turb}$ is the turbulent velocity dispersion. All these quantities are for the ambient medium. We used the values of $n\simeq0.29~\mathrm{cm^{-3}}$~\citep{lee1999}, $T \simeq 100\,\mathrm{K}$ and $\sigma_{\rm turb} \simeq 20~\mathrm{km\,s^{-1}}$~\citep{marasco2019}, and $B \simeq 6~\mu\mathrm{G}$ derived in Sect.~\ref{sec:equi_B}, and obtained $P_{e} \simeq 5.7\times10^{-12}\ \mathrm{dyn\,cm^{-2}}$. If the bubble is assumed to be spherical with a radius of 3~kpc the volume is $V=3.3\times10^{66}\,\rm cm^{3}$. Consequently, the total energy required to maintain the superbubble is $5/2\,P_e\,V\simeq1.9\times10^{55}$~erg.

At a height of 3~kpc, the wind speed is $\simeq400\,\rm km\,s^{-1}$ (see Fig.~\ref{fig:gal_wind}), implying a dynamical timescale of $\simeq7$~Myr for the supperbubble. This should be regarded as a lower limit, since the bubble’s expansion speed may be lower than the wind speed. Now we can check whether the star formation activity is enough to inflate the bubble over its dynamical time-scale. The mechanical luminosity of the wind can be estimated as $L_w=7\times10^{41}\epsilon\,(\rm SFR/M_\odot\,yr^{-1})\,erg\,s^{-1}$~\citep{veilleux2005}, where $\rm SFR$ is the star formation rate and $\epsilon$ is the thermalization efficiency. Using a $1.75~M_\odot\,\mathrm{yr}^{-1}$ SFR for NGC~3044 \citep{vargas2019} and adopting $\epsilon=0.5$ for star-forming galaxies~\citep{romero2018}, we estimate $L_w = 6\times10^{41}~\mathrm{erg\,s^{-1}}$. Hence, the
total energy injected over the course of 7~Myr is $1.3\times10^{56}\,\rm erg$, which would be easily sufficient to inflate the bubble.

Based on the observations in Fig.~\ref{fig:super_bubble}, the structure of the superbubble is likely due to the forward shock swept the ambient medium to form a thin shell containing warm gas traced by H$\alpha$ and H\textsc{i} emission~\citep[see][]{lee1997} and enhanced magnetic field and CREs traced by radio emission. The reverse shock heated the gas inside the bubble to about $10^6$~K to emit X-ray emission. This configuration is closely analogous to the large superbubble identified in the northern and southern halos of NGC~253~\citep{strickland2002,heesen_2009b,romero2018}.

For NGC~3044, higher resolution and sensitivity multi-wavelength observations are needed to confirm and explain the model of the superbubble. At present we only have morphological evidence from the radio continuum and the associated X-ray/$\rm H\alpha$ features, but not yet a kinematically confirmed. Future high-sensitivity H\textsc{i} observations, including the position–velocity diagrams across the structure, will be crucial to test this interpretation and to measure the expansion velocity of the shell.

\section{Conclusions}
\label{sec:conclu}

We re-processed the observations of NGC~3044 by ASKAP, improving the calibration and imaging, and obtained total intensity images at 943~MHz. We achieved an rms noise of 20~$\mu$Jy~beam$^{-1}$ and a resolution of $16\arcsec$. These images have a much better sensitivity than the previous images, allowing us to detect weak emission further away from the disk and thus study the propagation of CREs. 

Based on the high-resolution and high-sensitivity total intensity image from ASKAP, we examined the intensity profiles perpendicular to the disk. We found that intensity profiles can be fitted with two exponential components. We obtained synchrotron scale heights at 943~MHz. The mean values for the thin and thick disk components are \(h_{\mathrm{thin}} = 0.43\,\pm 0.13\,\mathrm{kpc}\) and \(h_{\mathrm{thick}} = 1.91\,\pm0.26\,\mathrm{kpc}\), respectively.
 
Using SPINNAKER, the advection and diffusion equations were solved to obtain vertical profiles of the synchrotron intensity at 943~MHz and 6~GHz and the spectral index. By comparing these profiles with observations, the propagation models of CREs can be determined. We found that CREs transport is advection-dominated. The advection speed, spectral index of injected CREs, and the scale heights of the magnetic field were also determined.  

The advection speed increases in a power law with $z$ and reaches the escape velocity of about 400~km~s$^{-1}$ at a height of 3~kpc. Beyond this height, cosmic rays can overcome the galactic gravitational potential and escape the galaxy. There is a good correspondence of multiwavelength emission extending from disk to halo, such as radio, X-ray, H\textsc{i}, and H$\alpha$, confirming the existence of a superwind causing the bulk motion of all the materials. 

At 943~MHz, we identify a possible superbubble with a radius $R\simeq3$~kpc in the northern halo, delineated by a bright H$\alpha$ rim and an H\textsc{i} shell, with its interior filled by hot soft X-ray emission. We estimate a lower limit to the energy required for the superbubble of $1.9\times10^{55}$~erg.

\begin{acknowledgments}
We would like to thank the critical comments by the reviewer which have significantly improved this paper. This research has been supported by the National SKA Program of China (2022SKA0120101, 2022SKA0120103). This work is also supported by the 16th Graduate Research and Innovation Project of Yunnan Province, China (Project No. KC-24249644). J.T.L. acknowledge the financial support from the National Science Foundation of China (NSFC) through the grants 12321003 and 12273111, the science research grants from the China Manned Space Program with grant no. CMS-CSST-2025-A04 and CMS-CSST-2025-A10, and Jiangsu Innovation and Entrepreneurship Talent Team Program through the grant JSSCTD202436. TW acknowledges financial support from the grant CEX2021-001131-S  funded by MICIU/AEI/ 10.13039/501100011033, from the coordination of the participation in SKA-SPAIN, funded by the Ministry of Science, Innovation and Universities (MICIU). This scientific work uses data obtained from Inyarrimanha Ilgari Bundara / the Murchison Radio-astronomy Observatory. We acknowledge the Wajarri Yamaji People as the Traditional Owners and native title holders of the Observatory site. The Australian SKA Pathfinder is part of the Australia Telescope National Facility, which is managed by CSIRO. Operation of ASKAP is funded by the Australian Government with support from the National Collaborative Research Infrastructure Strategy. Establishment of ASKAP, the Murchison Radio-astronomy Observatory and the Pawsey Supercomputing Centre are initiatives of the Australian Government, with support from the Government of Western Australia and the Science and Industry Endowment Fund.
\end{acknowledgments}

\facilities{ASKAP, MWA, VLA}
\software{ASKAPsoft~\citep{guzman2019}, CASA~\citep{mcmullin2007,casa2022}, \textsc{WSclean}~\citep{offringa2014}, Astropy~\citep{astropy2013,astropy2018,astropy2022}}

\bibliography{n3044}{}
\bibliographystyle{aasjournalv7}
\end{document}